\documentclass[11pt]{article}
\usepackage{subfig}
\usepackage{amssymb,amsmath,amsbsy}
\usepackage{mathrsfs}
\usepackage{dsfont}
\usepackage[top=1.2in, bottom=1.2in,left=0.9in,includefoot]{geometry}               % See geometry.pdf to learn the layout options. There are lots.
\usepackage{graphicx}
\usepackage{cite}
\newcommand{\newc}{\newcommand}
\newc{\ra}{\rightarrow}
\newc{\lra}{\leftrightarrow}
\newc{\beq}{\begin{equation}}
\newc{\eeq}{\end{equation}}
\newc{\barr}{\begin{eqnarray}}
\newc{\earr}{\end{eqnarray}}
\newc{\mdm}{M_{\mathrm{DM}}}
\newc{\aem}{\alpha_{\mathrm{em}}}
\newc{\adm}{\alpha_{\mathrm{DM}}}
\def\eq#1{eq.~(\ref{#1})}
\def\Eq#1{Eq.~(\ref{#1})}
\def\Eqs#1#2{Eqs.~(\ref{#1}) and (\ref{#2})}
\def\eqs#1#2{eqs.~(\ref{#1}) and (\ref{#2})}

\def\Ref#1{ref.~\cite{#1}}
\def\Rref#1{Ref.~\cite{#1}}

\def\Rrefs#1{Refs.~\cite{#1}}
\title{Searching for Secluded Dark Matter via Direct Detection of
Recoiling  Nuclei as well as Low Energy Electrons}
\author{A.~Dedes$^{\, a}$, I.~Giomataris$^{\, b}$,
K.~Suxho$^{\, a}$ and J.~D.~Vergados$^{\, a}$
\\[5mm]
$^{\, a}${\it Division of Theoretical Physics,
University of Ioannina,  GR 45110, Greece}
\\[3mm]
$^{\, b}${\it CEA, Saclay, IRFU, Gif-sur-Yvette, Cedex, France} }
\date{\today}                                           % Activate to display a given date or no date

\begin{document}

\maketitle

\begin{abstract}

Motivated by recent cosmic ray experimental results
there has been a proposition for a scenario where a secluded
dark matter particle annihilates, primarily, into Standard Model
leptons through a low mass mediator particle. 
We consider several varieties of this scenario
depending on the type of mixing among gauge
bosons and we study the
 implications   in novel direct dark matter experiments 
for detecting low energy recoiling electrons.
We find significant  event rates and time modulation effects,
especially in the case where the mediator is massless, that may be
complementary  to those from recoiling nuclei.

\end{abstract}

%%%%%%%%%%%%%%%%%%%%%%%%%%%%%%%%%%%
\section{Introduction}
\label{sec:intro}

The analysis of the  positrons excess (vs electrons) seen in cosmic ray spectra from PAMELA~\cite{Pamela1,Pamela2} in the energy region
above 10 GeV confirming previous results from HEAT~\cite{HEAT1,HEAT2}
and AMS-01~\cite{AMS01} experiments
together with very recent results from FERMI~\cite{FERMI}
and HESS~\cite{HESS} collaborations
 seems to suggest the presence of a WIMP that annihilates into leptons without any indication of annihilation into ($p, \bar{p}$) pairs or other hadrons (see \Rrefs{Strumia,Meade} for relevant analysis).
This is also reinforced by ATIC~\cite{ATIC} experiment which reports
excess of electron plus positron cosmic ray events in the energy region
$300 \lesssim E \lesssim 800$ GeV and also by
 signals from WMAP and EGRET \cite{WMAP1,WMAP2,EGRET} experiments. These phenomena can be explained by a scenario,
 originally proposed in \Ref{Holdom} - a
  subset of the so called {\em secluded Dark Matter} 
  scenarios~\cite{Maxim} -
  involving a
 new gauge boson $X_{\mu}$\footnote{In earlier 
 models~\cite{Finkbeiner} of   secluded 
 dark matter, WIMPs could be annihilated into new light scalar and gauge bosons.}, which couples to Standard Model  
 (SM) particles and the WIMP
 through kinetic vector boson mixing with the following 
 properties~\cite{models1} :
 %%%%%%%%%%%%%%%%%%%%%%%%%%%%
 \begin{equation}
 2 m_{e} \lesssim m_{X} \lesssim m_{\chi} \beta \lesssim m_{\chi} \adm
 \;,
 \label{wmc}
 \end{equation}
 %%%%%%%%%%%%%%%%%%%%%%%%%%%%
 where $m_{\chi} \beta$ is a typical non-relativistic WIMP momentum
and velocity $\beta \sim 10^{-3}$ 
inside the galactic halo and $\adm$ is the dark matter coupling. 
 It has been shown
 that if \eq{wmc} is satisfied then dark matter
 annihilation inside the halo to leptons 
 is enhanced by a Sommerfeld factor of
 $O(\adm/\beta)$~\cite{Sommerfeld} while
 annihilation to protons is simply kinematically forbidden. A typical 
 range of parameters that are going to be exploited in our analysis
 and satisfy \eq{wmc} are : 
 $m_{X} = 0.1-1$ GeV, $m_{\chi} = 0.1-1$ TeV
 and $\adm = \aem$. The new force mediated by the X-boson is
 a long range force indeed.
 We must note here that
 there is a choice of another viable
 possibility with an even lighter 
 mediator in MeV range that has been studied in \Ref{Boehm}.
 Our results for detecting low energy electrons are even more
 pronounced in this case. 
 
 There is also a possibility for the gauge boson mediator $X_{\mu}$ to
 couple to the SM gauge bosons through a mass mixing matrix 
 in a generalized gauge invariant way. These models are 
 frequently called St\"uckelberg models~\cite{Stueckelberg,Nath2}
 and are denoted as model type II in our classification. 
 A characteristic of these models 
is that the electromagnetic current couples to the 
dark sector 
through a massless pole identified as the physical photon.
As we shall see, this  
results in considerable and comparable
 rates in both nucleon or electron recoiling 
experiments.

Alternatively, it could be that there is a symmetry that renders dark matter particles leptophylic~\cite{Fox,Baek,Kribs,Yanagida,Ringwald}. 
This  symmetry is
spontaneously broken resulting in a massive gauge boson $X_{\mu}$
that couples directly to both leptons and  WIMP  at tree level. 
Again Sommerfeld enhancement dictates the mass of the $X$-boson
to be in the GeV (or sub GeV) range.
This is the  model III  that we consider in Chapter 2.

Within the three model categories mentioned above we want  :
\begin{enumerate}
\item to study the implications of this new force carrier on both traditional nucleon recoil, and untraditional electron recoil direct dark matter searches, and, 
\item to suggest new dark matter experiments involving the detection of electrons scattered by this carrier  providing a direct link to the
recently observed cosmic ray anomalous electron/positron events.
\end{enumerate}

So far there is a dedicated  analysis for electron recoils in DAMA
experiment~\cite{Bernabei}
 with energies approximately 5 KeV. 
Our analysis  investigates
recoiling electrons with 
energies as low as 10 eV, and suggests an experimental method on
how to reach such low energies. It is therefore  complementary
to the analysis of~\Rref{Bernabei}.

The structure of this article is as follows :  In section 2 we present
a field theory setup which helps to categorize   three representative
model examples that have recently been studied in detail.   In 
section 3,   we present event rate predictions
for conventional nucleon recoil detection for the models studied.
In section 4,  we deal 
with the not so familiar  methods of electron recoil detection  
rates together with time modulation effects. We also make
a proposition of   a prototype  experiment  to be exploited
in discovering low energy  electrons ejected from WIMP + atom
collisions. In section 5 we present our conclusions.

\section{Theory Setup and Model Categories}
\label{sec:models}
%\subsection{Propagator Mixing Formalism}

In this section,
we formulate the problem of the Standard Model coupled to, for
simplicity, an Abelian
dark sector with arbitrary kinetic or mass mixing terms allowed by
Lorentz, gauge symmetries and renormalizability.
Our formulae  are then applied in subsequent sections to
make predictions  for  event rates in
dark matter detection experiments.

To read out the gauge boson  propagators
we start by writing  the general renormalizable  form
of the Lagrangian :
%%%%%%%%%%%%%%%%%%%%%%%%%%%%%%%%%%%
\begin{eqnarray}
\mathscr{L} \ = \ -\frac{1}{4}\: {\bf \Phi}^{T}_{\mu\nu} \, \mathcal{K} \, {\bf \Phi}^{\mu\nu}
\ + \ \frac{1}{2} \: {\bf \Phi}_{\mu}^{T} \,\mathcal{M}^{2} \:  \, {\bf \Phi}^{\mu}
\ -\  \frac{1}{2} \: \partial^{\mu} {\bf \Phi}_{\mu}^{T} \, \mathrm{\Xi} \,
\partial^{\nu} {\bf \Phi}_{\nu} \ + \ {\bf J_{\mu}}^{T} \, {\bf \Phi}^{\mu}
 \;, \label{lag}
\end{eqnarray}
%%%%%%%%%%%%%%%%%%%%%%%%%%%%%%%
where ${\bf \Phi}_{\mu\nu} = (\partial_{\mu} {\bf \Phi}_{\nu} -
\partial_{\nu} {\bf \Phi}_{\mu})$ is a $N$-column matrix  field
strength tensor corresponding to a N-column ${\bf \Phi}_{\mu}$ vector field,
 ``${T}$'' denotes the transpose of a matrix,
 $\mathcal{K}$ and  $\mathcal{M}^{2}$ are  real
and symmetric $N\times N$ matrices with model
dependent elements to be specified below
and $\mathrm{\Xi}$ is the gauge fixing $N\times N$ symmetric matrix necessary
to remove unphysical gauge degrees of freedom.
Interaction terms are encoded in the last term of \eq{lag}
where an external current ${\bf J_{\mu}}$ associated with symmetries, couples to the gauge fields.

One has to notice that  elements of the mass matrix $\mathcal{M}^{2}$
should be further restricted by electromagnetic gauge invariance.
Phenomenologically speaking, there should always be a pole on the propagator
$\langle {\bf \Phi}_{\mu} {\bf \Phi}_{\nu} \rangle$   corresponding to the
massless photon i.e., the determinant of the inverse propagator at zero
momentum must be exactly zero. Furthermore, without loss of generality,
we can always assume that the diagonal elements of $\mathcal{K}$ are
normalized to unity.

It is standard textbook exercise to find the Feynman propagator,
$\widetilde{\mathcal{D}}_{\mu\nu}(p)$ with momentum $p$, for the gauge field ${\bf \Phi}^{\mu}$
which in  momentum space reads,
%%%%%%%%%%
\begin{eqnarray}
i \, \widetilde{\mathcal{D}}_{\mu\nu}(p)\ = \ (\mathcal{K}\:p^{2} - \mathcal{M}^{2} )^{-1}
\, \left(g_{\mu\nu}  - \frac{p_{\mu}p_{\nu}}{p^{2}} \right ) \ + \
\left(\mathrm{\Xi}\, p^{2} - \mathcal{M}^{2} \right )^{-1} \,
 \frac{p_{\mu}p_{\nu}}{p^{2}} \;.\label{prop}
\end{eqnarray}
%%%%%%%%%%%%%%%
At lowest order in $\hbar$, interactions among fields are stored in the action functional
%%%%%%%%%%%%%%
\begin{eqnarray}
S[\widetilde{{\bf J}}] \ = \ \frac{1}{2}\: \int \frac{d^{4}p}{(2\pi)^{4}} \;
\widetilde{\bf J}_{\mu}^{T}(p) \;  [ i\widetilde{\mathcal{D}}^{\mu\nu}(p) ] \;
\widetilde{\bf J}_{\nu}(-p) \;,
\label{int}
\end{eqnarray}
%%%%%%%%%%%%%%%%%%
where $\widetilde{\bf J}_{\mu}(p)$ is the vector current in momentum space.
\Eqs{prop}{int} are what we actually need to describe observables
that arise from  mixing  dark (or hidden)  and visible
gauge bosons. As a simple example, consider the electromagnetic and the dark
gauge boson current. Then  in \eq{lag}, it is ${\bf J}^{T}_{\mu} = (e J_{\mu}^{\rm e.m}\, , \, g_{X} J_{\mu}^{\rm dark})^{T}$. It is then clear from \eq{int} that
interactions between the visible and the dark sector will involve
off diagonal elements of the propagator (\ref{prop}). Observables,
like nucleon recoil event rates  can  easily be described using the
above propagator mixing formalism~\cite{Sakurai},
by simply finding the inverse  matrices such in \eq{prop} for
a given model.
We remark here that the propagator mixing formalism works equally
well in different current basis such as $Q-T_{3}$ or $Y-T_{3}$.

\subsection{Model I : Non-standard Kinetic Mixing $\mathcal{K}$}

Models in this category~\cite{Holdom,Maxim}
have been recently exploited in \Ref{models1}
as candidates for explaining positron 
excess in cosmic ray
data experiments. In its simplest form, the dark matter particle,
$\chi$,  is charged
under a `dark' $U(1)_{X}$ and the corresponding `dark' gauge boson
$X_{\mu}$ mixes with
the photon $A_{\mu}$ and Z-gauge boson, $Z_{\mu}$.
 Annihilations of dark matter particles into {\it only} SM leptons
(and not quarks) are
kinematically allowed
when the intermediate gauge boson has a mass at the GeV
scale.

In notation of \Ref{Cheung} and in basis
$(A_{\mu}, X_{\mu}, Z_{\mu})$ (or else $Q-T_{3}$) our matrices
$\mathcal{K}$ and $\mathcal{M}^{2}$ appeared in \eq{prop},  become:
%%%%%%%%%%%%%%%%%%%%%%%%
\begin{eqnarray}
\mathcal{K} \ = \  \left( \begin{array}{ccc}
                           1  &  -\epsilon \cos\theta_{W}  & 0 \\
                           -\epsilon \cos\theta_{W} & 1 & \epsilon \sin\theta_{W} \\
                           0 & \epsilon \sin\theta_{W} & 1 \end{array} \right)
                           \quad , \quad
\mathcal{M}^{2} \ =\  \left( \begin{array}{ccc}
                                  0 & 0 & 0 \\
                                  0 & m_{X}^{2} & 0 \\
                                  0 & 0 & m_{Z}^{2} \end{array} \right ) \;, \label{mat1}
\end{eqnarray}
%%%%%%%%%%%%%%%%%%%%%%%
where $m_{X}$ is the mass of the
exotic  gauge boson, $m_{Z}$ is the mass of
Z-boson, $\theta_{W}$ is the weak mixing angle and $\epsilon$ is a small ($\approx 10^{-3}$) mixing
parameter between $U(1)_{Y}$ and $U(1)_{X}$  field strength tensors.
%%%%%%%%%%%%%%%%%%%%%%%%%%%%%
\begin{figure}[t]
\begin{center}
\includegraphics[height=20mm]{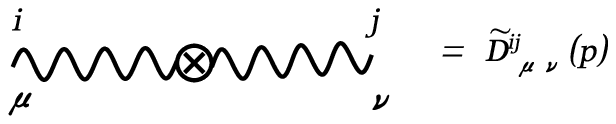}
\caption{\sl Diagramatic form of Feynman propagator appeared in  \eq{prop} between gauge boson ``flavours'' $i$ and $j$.
For explicit expressions in model~I see
  Eqs.(\ref{gg})-(\ref{ZZ}); for model~II see \eq{eff2}.}
\label{fig1}
\end{center}
\end{figure}
%%%%%%%%%%%%%%%%%%%%%%%%%%%%%%
Working in Feynman gauge ($\Xi = \mathbf{1}_{3\times 3}$) and
keeping  up to $\epsilon^{2}$-terms it is easy
to work out the mixed
propagators  $\widetilde{\mathcal{D}}^{ij}_{\mu\nu}(p)$,
depicted in Fig.\ref{fig1}, between
 photon,  $X$ and $Z$-gauge bosons, labeled 1,2,3, respectively :
%%%%%%%%%%%%%%%%%%
\begin{eqnarray} \label{gg}
i \, \widetilde{\mathcal{D}}^{11}_{\mu\nu}(p)\ &=& \
\frac{g_{\mu\nu}}{p^{2}} \ + \ \frac{\epsilon^{2}\, \cos^{2}\theta_{W}}{p^{2} - m_{X}^{2}}
\, \left(g_{\mu\nu}  - \frac{p_{\mu}p_{\nu}}{p^{2}} \right )
\ + \ O(\epsilon^{3}) \;, \\[3mm]
%%%%%%%%%%%%%%%%%%%%%%%%%%%
i \, \widetilde{\mathcal{D}}^{12}_{\mu\nu}(p)\ &=& \
\frac{\epsilon \cos\theta_{W}}{p^{2}-m_{X}^{2}}
\, \left(g_{\mu\nu}  - \frac{p_{\mu}p_{\nu}}{p^{2}} \right )
\ + \ O(\epsilon^{3}) \;, \\[3mm]\label{gb}
%%%%%%%%%%%%%%%%%%%%%%%%%%%
i \, \widetilde{\mathcal{D}}^{13}_{\mu\nu}(p)\ &=& \
-\frac{\epsilon^{2} \, p^{2} \, \cos\theta_{W} \sin\theta_{W}}{(p^{2} - m_{X}^{2})
(p^{2} - m_{Z}^{2})}
\, \left(g_{\mu\nu}  - \frac{p_{\mu}p_{\nu}}{p^{2}} \right )
\ + \ O(\epsilon^{3}) \;, \\[3mm]\label{gZ}
%%%%%%%%%%%%%%%%%%%%%%%%%%%%
i \, \widetilde{\mathcal{D}}^{22}_{\mu\nu}(p)\ &=& \
\frac{g_{\mu\nu}}{p^{2} - m_{X}^{2}} \ + \
\frac{\epsilon^{2}\, p^{2}\: (p^{2} - \cos^{2}\theta_{W} m_{Z}^{2})}{(p^{2} - m_{X}^{2})^{2} (p^{2} - m_{Z}^{2})}
\, \left(g_{\mu\nu}  - \frac{p_{\mu}p_{\nu}}{p^{2}} \right )
\ + \ O(\epsilon^{3}) \;, \\[3mm]\label{bb}
%%%%%%%%%%%%%%%%%%%%%%%%%%%%%
i \, \widetilde{\mathcal{D}}^{23}_{\mu\nu}(p)\ &=& \
-\frac{\epsilon \, p^{2} \sin\theta_{W}}{(p^{2}-m_{X}^{2}) \, (p^{2}-m_{Z}^{2})}
\, \left(g_{\mu\nu}  - \frac{p_{\mu}p_{\nu}}{p^{2}} \right )
\ + \ O(\epsilon^{3}) \;,  \\[3mm]\label{bZ}
%%%%%%%%%%%%%%%%%%%%%%%%%%%%%%
i \, \widetilde{\mathcal{D}}^{33}_{\mu\nu}(p)\ &=& \
\frac{g_{\mu\nu}}{p^{2} - m_{Z}^{2}} \ + \
\frac{\epsilon^{2} \, p^{4}\: \sin^{2}\theta_{W}}{(p^{2} - m_{X}^{2}) (p^{2} - m_{Z}^{2})^{2}}
\, \left(g_{\mu\nu}  - \frac{p_{\mu}p_{\nu}}{p^{2}} \right )
\ + \ O(\epsilon^{3}) \;.  \label{ZZ}
\end{eqnarray}
%%%%%%%%%%%%%%
Some remarks are in order : {\it i)} among the three physical masses only
$m^{2}_{X}$ mass is shifted by an amount of $m_{X}^{2}\epsilon^{2}$
that we ignore {\it ii)} gauge invariance for the off diagonal
propagator terms is
preserved as should be the case.
As far as the effective action \eq{int} is concerned,
additional statements are in order:
%%%%%%%%%%%%%%%%%%%%%%%%%%
\begin{itemize}
\item The single pole $[1/p^{2}]$
appears only in $J_{\rm e.m}\cdot J_{\rm e.m}$ exchange
as usual in the SM.
\item A pole $[1/(p^{2}-m_{X}^{2})]$
for the exotic boson $X_{\mu}$ appears, apart from
$J_{X}\cdot J_{X}$ exchange, also in $J_{\rm em}\cdot J_{X}$
exchange at $O(\epsilon)$.
\item There is exchange of current $J_{X}\cdot J_{Z}$ i.e., neutrinos and
dark matter particles, through a double pole of $X$ and $Z$
at order $\epsilon$.
\item  There is exchange of $J_{\rm em}\cdot J_{Z}$
at order $\epsilon^{2}$
via double pole of $X$ and $Z$
\end{itemize}
%%%%%%%%%%%%%%%%%%%%%
The $\epsilon \approx 10^{-3}$-term in the kinetic mixing
can naturally arise as a result of mixing two
 $U(1)$'s at high energies - a mechanism that  it was first proposed in
 \Rref{Holdom}. Furthermore, $X$-boson contributions
 to the muon anomalous magnetic moment relative to the
 SM expectation, $\Delta \alpha_{\mu}=
 \alpha_{\mu}^{\mathrm{exp}} - \alpha_{\mu}^{\mathrm{SM}} =
 (290 \pm 90) \times 10^{-11}$~\cite{HAGI}, are easily found using
 \eq{gg} to be 
 %%%%%%%%%%
 \begin{equation}
 \Delta \alpha_{\mu} \ = \ \frac{\aem}{3\pi} \: \epsilon^{2}
 \cos^{2}\theta_{W}\: \biggl (\frac{m_{\mu}}{m_{X}} \biggr )^{2}\;,
 \qquad \mathrm{for}\quad  \frac{m_{\mu}}{m_{X}} \ll 1 \;.
 \label{dam1}
 \end{equation}
 %%%%%%%%%%%%%%
 This  requires
 $\epsilon \lesssim 3\times 10^{-2}$ for $m_{X} \simeq 1$ GeV
  where the equality accounts for the 2$\sigma$ upper limit on
  $\Delta \alpha_{\mu}$.
Of course there are many other constraints on the mixing parameter
 $\epsilon$ from direct or indirect collider searches and we refer
 the reader to  recent work in \Rrefs{Maxim2,Zurek,Bjorken,Maxim3}. 
  For example,
 as we see from eqs. (\ref{gg}), (\ref{gZ}) and (\ref{ZZ}) corrections to oblique 
 electroweak  observables arise at order $\epsilon^{2}$ similar to
 the case of muon anomalous magnetic moment.

\subsection{Model II : Non-standard Mass Mixing, $\mathcal{M}^{2}$}
\label{sec:mod2}

Models belonging to this category are usually referred to as Stueckelberg
models~\cite{Stueckelberg}. A recent account on  ``Stueckelberg'' extensions
of the Standard Model can be found in \Rref{Nath}.
Here, it is more convenient to work on $Y-Y_{X}-T_{3}$ basis
$(B_{\mu}, X_{\mu}, A_{\mu}^{3})$. We now assume  that only
the matrix $\mathcal{M}^{2}$ is nontrivial,
%%%%%%%%%%%%%%%%%%%%%%%%%%
\begin{eqnarray}
\mathcal{K} \ = \  \left( \begin{array}{ccc}
                           1  &  0 & 0 \\
                           0 & 1 & 0 \\
                           0 & 0 & 1 \end{array} \right)
                            \;, \qquad
\mathcal{M}^{2} \ =\  \left( \begin{array}{ccc}
\frac{1}{4} \: g^{2}_{Y} v^{2} + m_{Y}^{2} & m_{Y} \, m_{X} & -\frac{1}{4} \: g_{Y} g \: v^{2} \\
m_{Y}\, m_{X} & m_{X}^{2} & 0 \\
 -\frac{1}{4} \: g_{Y} g \: v^{2} & 0 &  \frac{1}{4}\: g^{2}
  v^{2} \end{array} \right ) \;, \label{mat2}
\end{eqnarray}
%%%%%%%%%%%%%%%%%%%%%%%
where $g_{Y},g$ are the $U(1)_{Y}, SU(2)_{L}$ gauge couplings respectively,
$m_{Y}^{2}$ is a mass term for the hypercharge gauge field $B_{\mu}$
and $v$ is the vacuum expectation value. The form of the upper
left $2\times 2$  $\mathcal{M}^{2}$
matrix guarantees electromagnetic gauge invariance
i.e., massless photon. Furthermore,
the zero elements (23) and (32)  guarantee that
neutrinos are not charged under electromagnetism.
Demanding that the inverse propagator has poles at
the physical masses, $\det[p^{2} - \mathcal{M}^{2}]|_{p^{2}=m_{i}^{2}}=0$
where $m_{i}=0,m_{X},m_{Z},$ we find that
the photon mass is zero to all orders in $m_{Y}$,
the dark gauge boson and the Z-boson masses
are not altered up to $O(m_{Y}^{2})$, and thus
$m_{Z}^{2} = \frac{1}{4}(g^{2} + g_{Y}^{2}) v^{2} + O(m_{Y}^{2})$.

Following  \eq{int} we obtain the following effective action,
%%%%%%%%%%%%%%%%%%%%%%%%%%%
\begin{eqnarray}
S[J] \ &=& \ \frac{1}{2}\, \int \frac{d^{4}p}{(2\pi)^{4}} \  \left \{
\left [ e^{2} \: J_{\rm e.m}(p) \cdot J_{\rm e.m}(-p) - 2\: e^{2} \: \frac{g_{X}}{g_{Y}} \frac{m_{Y}}{m_{X}}\, J_{\rm e.m}(p) \cdot J_{X}(-p)
\right ]\, \frac{1}{p^{2}}  \right. \nonumber  \\[3mm]
  &+& \left. \left[g_{X}^{2}\: J_{X}(p) \cdot J_{X}(-p)\:
  \left (1-\frac{m_{X}^{2}}{m_{Z}^{2}}\right )
   \ + \ 2\: e^{2} \frac{g_{X}}{g_{Y}}\frac{m_{Y}}{m_{X}}\:
  J_{\rm e.m}(p)\cdot J_{X}(-p) \right. \right. \nonumber \\[3mm]
   &-& \left. \left. 2\: g_{Y} g_{X} \frac{m_{X}\: m_{Y}}{m_{Z}^{2}}\:
    J_{X}(p)\cdot J_{Y}(-p)
  \right]\: \frac{1}{p^{2}-m_{X}^{2}} \right. \nonumber \\[3mm]
   &+& \left. \left [ g^{2}\: J_{Z}(p) \cdot J_{Z}(-p) \  + \
  g_{X}^{2} \frac{m_{X}^{2}}{m_{Z}^{2}} J_{X}(p)\cdot J_{X}(-p)
  \right. \right. \nonumber \\[3mm]
  &+&  \left. \left. 2 g_{Y} g_{X} \frac{m_{X} \: m_{Y}}{m_{Z}^{2}} \: J_{X}(p)\cdot J_{Y}(-p) \right ]\frac{1}{p^{2}-m_{Z}^{2}}
  \right \}  \ + \ O(m_{Y}^{2})
   \;, \label{eff2}
\end{eqnarray}
%%%%%%%%%%%%%%%%%%%%%%%%%%%
where $e \equiv g_{Y} g/\sqrt{g_{Y}^{2} + g^{2}}$ is the electron charge.
Furthermore,   $J_{\rm e.m}(p) = J_{A_{3}}(p) + J_{Y}(p)$ is the
momentum space Fourier transform of the
electromagnetic current, i.e.,  $J^{\mu}_{\rm e.m} =
\sum_{f}  Q_{f} \bar{f} \gamma^{\mu} f $ with $Q_{f} e $ being the charge of
a generic fermion $f$. The dark current $J_{X}$ obtains an analogous
formula with obvious replacement of charge $Q_{f}  e$ by another
(hyper)charge,
 $Q_{X}$. Of course, if fermions under consideration
 are Majorana particles then the
 corresponding current has only axial-vector form.
In addition, $J_{Z}$ denotes the Fourier transform of the
Standard Model neutral current $J_{Z}^{\mu} =
\frac{1}{\cos\theta_{w}} (J^{\mu}_{A_{3}} - \sin^{2}\theta_{w} \: J^{\mu}_{\rm e.m})$ where the electromagnetic current is, as usual in the SM, the
sum of the third component of the isospin $J^{\mu}_{A_{3}}$
and hypercharge currents $J^{\mu}_{Y}$.
The physics of \eq{eff2} is now transparent : to order O($m_{Y}$),
 there are interactions
between the electromagnetic  $J_{\rm e.m}$  and dark current
$J_{X}$ mediated by the photon {\it i.e.,} the dark matter particle is charged,  and interactions between the
hypercharge $J_{Y}$ and dark current $J_{X}$ mediated by ($X$ or
$Z$) gauge bosons, respectively. An estimate of the dominant
contribution to $\Delta \alpha_{\mu}$ results in  an upper bound
$\frac{m_{Y}}{m_{X}} \lesssim 9\times10^{-4}$, where 
a 2$\sigma$ bound on 
$\Delta \alpha_{\mu}$  is taken from \Rref{HAGI}.

\subsection{Model III : Direct coupling, no mixing}
\label{sec:mod3}

In this model, some of the SM leptons (but not quarks) 
$\ell_{L},e_{R}$ and
the WIMP particle $\chi$ are coupled
directly to the dark gauge boson $X_{\mu}$ in principle with different
couplings\footnote{Various possibilities on how this is realized
can be found in~\Rref{Fox}.} :
%%%%%%%%%%%%%%%%%%%%%%%
\begin{eqnarray}
J_{X}^{\mu} \ = \ g' \: Y'(e_{L}) \, \bar{\ell}_{L} \gamma^{\mu} \ell_{L}
\ + \  g' \: Y'(e_{R}) \, \bar{e}_{R} \gamma^{\mu} e_{R} \ + \ g_{X}
\: Y'(\chi)\, \bar{\chi} \gamma^{\mu} \chi \;, \label{Jxmu}
\end{eqnarray}
%%%%%%%%%%%%%%%%%%%%%%%%
where $Y'(e_{L},e_{R})=(1,-1)$
denotes the particle hypercharge under the new
gauge symmetry.
As it has been suggested in \Rref{Strumia,Fox,Baek,Yanagida,Ringwald},
this could be an anomaly free gauged  $U(1)_{L_{e}-L_{\tau}}$.
Of course a new Dirac fermion $\chi$ would be playing the role
of dark matter particle is also gauged
 under this symmetry with $Y'(\chi)=1$.
Because we have already discussed the effects of the kinetic and
mass mixing in the previous models, without loss of generality,
we assume that these mixing matrices are trivial 
in this model at tree level\footnote{Of course mixing of the 
$X_{\mu}$ gauge boson with the $U(1)_{Y}$ is inevitable 
at one loop. Its magnitude is calculable : 
$\epsilon \simeq \alpha^{'2} \log\frac{m_{\tau}}{m_{\mu}} 
= 2 \times 10^{-4}$
for $\alpha' = \aem$. All the rest will then proceed following
eqs.(\ref{gg} - \ref{ZZ}) of model I.}.
If $X_{\mu}$ does not couple to the muon then the most 
important constraint on $\alpha'={g'}^{2}/4\pi$ will arise from the $\nu - e$
scattering at low $q^{2}$ :
\beq 
\frac{\alpha'}{m_{X}^{2}}  \ \lesssim \ 7 \times 10^{-7}\;.
\label{enu}
 \eeq
 We shall use this bound when discussing electron 
 recoil detection rates in section~\ref{sec:unco} as is typically
 comparable (most of the time better) 
 with other direct experimental bounds arising from LEP
 or meson factories. 
 If the $X_{\mu}$ vector boson couples to electrons and muons 
 instead then there is a comparable bound to \eq{enu} from
 the muon anomalous magnetic moment. Following 
$\Delta \alpha_{\mu} =\frac{\alpha'}{3\pi}\frac{
m^{2}_{\mu}}{m^{2}_{X}}$ for $m_{\mu} \ll m_{X}$,
 there is a bound
 %%%%%%%%%%%%%%%%%%
 \beq
 \frac{\alpha'}{m_{X}^{2}} \ \lesssim \ 4.4 \times 10^{-6} \;.
 \label{bound-III}
 \eeq
 %%%%%%%%%%%%%%%%%%%%%

\section{Conventional WIMP searches}
\label{sec:con}

Conventional DM searches deal with phenomena of WIMPs scattered
of a nucleus. The study of the recoil energy spectrum is the
primary goal of experiments such as CDMS~\cite{CDMS},
XENON~\cite{XENON} and DAMA~\cite{DAMA}.
For models we described in the previous section there are two
cases which have been discussed recently in the literature that
could explain the recent anomalous  cosmic ray events:
\begin{itemize}
\item[a)] The lightest mediator is massless and
\item[b)] the lightest  mediator is massive with mass around the proton mass ($m_{p}$),
\end{itemize}
in addition to the assumption that
%%%%%%%%%%%%
\beq
 m_{p} \  \ll \  m_{\chi} \;,
\label{assum}
\eeq
%%%%%%%%%%%%%
where $m_{\chi}$ is the WIMP mass.
Only model II belongs to the first category and models I,II  belong to the second since by definition, there is no direct 
coupling of X-boson to quarks in model III [see however footnote 2].
 In the following subsections
we present the
WIMP-nucleon cross section for both cases (a) and (b).

\subsection{Massless Mediator}
\label{mm1}

%%%%%%%%%%%%%%%%%%%%%%%%%%%%%%%%%%%
\begin{figure}[t]
   \centering
   \includegraphics{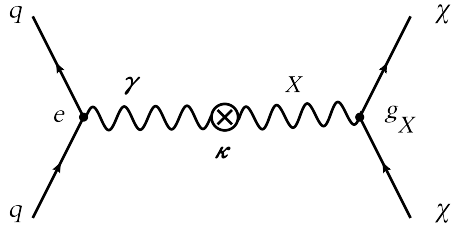} % requires the graphicx package
   \caption{\sl A Feynman diagram leading to the direct interaction  of the WIMP $\chi$ to the quarks relevant for direct detection of dark matter. The process is mediated by the physical photon. The cross indicates merely that the exotic gauge boson has a small admixture of the photon. Similarly the WIMP  can also couple to electrons.}
 \label{Fig:B-photon}
  \end{figure}
%%%%%%%%%%%%%%%%%%%%%%%%%%%%%%%%%%%%%%%
The differential WIMP-proton cross section in the rest frame of the initial proton is given by:
%%%%%%%%%%%%%%%%%%%%%%%%%
\beq
d\sigma \ = \ \frac{s(\beta)}{\beta}\:
\frac{e^{2}\left({g_{X} \kappa}\right)^2}{q^4}\:
\frac{d^3{\bf p'}}{(2 \pi)^3}\, \frac{d^3{\bf q} }{(2 \pi)^3}\:
 (2 \pi)^3\: \delta^{(3)}\: ({\bf{p-p'-q}})\: (2 \pi)\: \delta(T-T{'}-T_q)\;.
 \label{eq:3.16}
\eeq
%%%%%%%%%%%%%%%%%%%%
In the above equation   $\bf{p',p}$  are the momenta of the initial WIMP and the final WIMP and  $\bf{q}$ the momentum transfer to the nucleon and $T=p^2/2 m_{\chi},~T'=(p')^2/2 m_{\chi}$ and $T_q=q^2/2 m_p$, are respectively the corresponding kinetic energies in the non relativistic limit.
Furthermore,  $\beta$ is the WIMP velocity and $s(\beta)=1$ for a WIMP which is a Dirac fermion, while  $s(\beta)=\beta^2$ in case it is Majorana one~\cite{Vergados}.\footnote{The Majorana fermion does not possess electromagnetic properties. Hence only the $\gamma_{\mu}\gamma_5$ of the WIMP --$X$-boson interaction contributes.}
One finds that the momentum transfer and the final nucleon energy are given by:
%%%%%%%%%%%%%%%%%%%%%
\beq
q=2 \mu_r \upsilon \xi \approx2 m_p \upsilon \xi~~,~~ T_q \approx2 m_p\upsilon^2 \xi^2 \;,
\label{Eq1:q-tranfer}
\eeq
%%%%%%%%%%%%%%%%%
where $\mu_r$ is the WIMP-nucleon reduced mass, $m_{p}$ is the
proton mass and $0 \leq\xi\leq 1$ is
the cosine of the 
 angle between the incoming WIMP and the outgoing nucleon.
Integrating over the momentum of the outgoing WIMP and the magnitude of the momentum of the final hadron as well as the $\phi$-angle  one
finds :
%%%%%%%%%%%%%%%%
\beq
d\sigma \ = \ \frac{s(\beta)}{\beta}\:
\frac{e^{2}\left({g_{X} \kappa}\right)^2}{2 \pi}\:
\frac{1}{(2m_p)^2} \: \frac{d\xi}{\upsilon^3 \xi^3}\;.
\label{ds}
\eeq
%%%%%%%%%%%%%%%%
The above expression exhibits, of course, the infrared divergence. 
We will impose a low momentum cut off $E_{\mathrm{th}}/A$ provided 
 by the energy threshold $E_{\mathrm{th}}$, 
 where A is the mass number of the target, i.e.
%%%%%%%%%%%%%%%%
\beq
\xi_\mathrm{min}=\sqrt{\frac{E_\mathrm{th}}{(2A m_p \beta^2)}}\;.
\eeq
%%%%%%%%%%%%%
%\subsection{Contact with the Direct Dark Matter Searches}
Thus the total cross-section for a Majorana WIMP is given by:
%%%%%%%%%%%%%%%%%%%%%
\beq
\sigma \ = \
 \frac{ \alpha}{2} \left( {g_{X} \kappa}\right)^2 
 \frac{1}{(m_p)^2} \left ( \frac{Am_p}{E_\mathrm{th}}
 - \frac{m_{p}}{T_{\rm max}}\right )
 \approx \frac{ \alpha}{2} \left( {g_{X} \kappa}\right)^2\frac{1}{(m_p)^2}
 \frac{Am_p}{E_\mathrm{th}} \;.\label{eq:3.20}
\eeq
%%%%%%%%%%%%%%%%%%%%%%
\Eq{eq:3.20} shows a much stronger dependence of the event rate
on the threshold energy $E_\mathrm{th}$ due to the adopted cut-off
$E_\mathrm{cut-off}=E_\mathrm{th}/A$.
It is interesting to note that this cross section is independent of the WIMP velocity (in the case of a Dirac WIMP the  extracted from the data cross section must be  multiplied by  $\beta^2$).
We distinguish two cases :

\begin{enumerate}
\item The case of Majorana WIMP. We   find:
\beq
\sigma \approx 1.6 \times 10^{-30} \mbox{ cm}^{2}  \left( { g_{X} \kappa}\right)^2   \frac{2Am_p}{E_{th}}\:.
\eeq
 The direct dark matter experiments have recently set on the coherent nucleon cross section the limits:
\begin{itemize}
\item The CDMSII experiment \cite{CDMS}:\\
The best limit is $6.6\times 10^{-44}$ $\mathrm{cm}^{2}$. The extracted value depends, however, on the assumed WIMP mass. So it can vary between $ 6.6\times 10^{-44}$ and $ 6.6\times 10^{-42}$ 
$\mathrm{cm}^{2}$.
\item The XENON10 collaboration \cite{XENON}\\
They extract $8.8\times 10^{-44} $ $\mathrm{cm}^{2}$ 
and $4.5\times 10^{-44} $ 
$\mathrm{cm}^{2}$ for WIMP masses of 100 and 30 GeV respectively.
\end{itemize}
For our purposes we will assume that the extracted from the data  nucleon cross section is $10^{-7}$pb = $10^{-43}$cm$^2$. Furthermore we will take as a reference  a threshold energy of $5.0$ KeV and examine the sensitivity of our results to the experimental threshold.
Using the experimental limit,  $\sigma_p\leq 1.0 \times 10^{-43}$ 
$\mathrm{cm}^{2}$,
 we can write:
\beq
\frac{\mbox{Rate(new)}}{\mbox{Rate(conventional)}}=1.6\times 10^{6} \frac{Z^2}{A^2}\left( { g_{X} \kappa}\right)^2   \frac{Am_p}{E_{th}}\;.
\eeq
Note that the coherence factor now is $Z^2$, since in the case of the photon only the protons of the target contribute. Adopting a threshold value of 5 KeV, we get
\beq
\frac{\mbox{Rate(new)}}{\mbox{Rate(conventional)}}=3.0\times 10^{18} \frac{Z^2}{A}\left( { g_{X} \kappa}\right)^2\;.
\eeq
For the Ge target $(A=73,Z=32)$ we get
\beq
\frac{\mbox{Rate(new)}}{\mbox{Rate(conventional)}}=4.3\times 10^{19} \left( { g_{X} \kappa}\right)^2\;,
\eeq
which leads to the limit:
\beq
| g_{X} \kappa| \leq \sqrt{ \frac{1}{0.43 \times 10^{19}}}= 1.6 \times 10^{-10}\;.
\eeq
%%%%%%%%%%%%%%%%%%%%%%%
From the second term in \eq{eff2} and assuming
that $\adm = g_{X}^{2}/4\pi = \aem$
one can easily translate this into bounds on
the model II parameters for Majorana WIMP  :
%%%%%%%%%%%%%%%%%
\beq
Q_{X}\: \frac{m_{Y}}{m_{X}} \lesssim
0.54\times 10^{-10}\;, \qquad \mathrm{model ~II}
\;.\label{bound1}
\eeq
\item The case of a Dirac WIMP. We find:
\beq
\sigma \approx \frac{1}{\beta^2}\frac{ \alpha}{2} \frac{1}{(m_p)^2}\left( { g_{X} \kappa}\right)^2   \frac{Am_p}{E_{th}}\;.
\eeq
If we knew the coupling $| g_{X} \kappa|$  we could incorporate this into the evaluation of the nuclear cross section, fold it with the velocity distribution and proceed with the evaluation of the event rate. Since, however, we like to constrain the parameter $|g_{X} \kappa|$ we will employ an average velocity:
\beq
\sigma \rightarrow < \sigma > \ \approx \ < \frac{1}{\beta^2} 
>\frac{ \alpha}{2} \frac{1}{(m_p)^2}\left( {g_{X} \kappa}\right)^2   \frac{Am_p}{E_{th}}\;.
\eeq
But for a Maxwell - Boltzmann distribution {\it i.e., }
$
<\frac{1}{\beta^2}> \ \rightarrow \ \frac{3}{< \beta^2>}
$,
we obtain the constraint:
\beq
|g_{X} \kappa| \leq 1.6 \times 10^{-10}\frac{\sqrt{<\beta^2>}}{\sqrt{3}}\approx0.8\times 10^{-13}\;,
\label{nc1}
\eeq
from which the bound on model II for $\adm = \aem$,
\beq
Q_{X}\: \frac{m_{Y}}{m_{X}} \lesssim
 0.27\times 10^{-13}\;, \qquad \mathrm{model ~II}
\;,\label{bounddirac}
\eeq
is found. As expected the limit is now more stringent than in \eq{bound1}.
\end{enumerate}
%%%%%%%%%%%%%%
 The results for the Xe target are similar.
This bound  is  by many orders of magnitude stronger than
the one obtained from electroweak fits~\cite{Nath} or $(g-2)_{\mu}$
[see discussion towards the end of section~(\ref{sec:mod2})].
The corresponding bound for
Dirac WIMP is about three orders of magnitude more stringent.
This means that additional mechanisms should be added in model II
(St\"uckelberg type of \Rref{Nath} for example) in order  to
efficiently depleting the WIMP in the early universe (the diagram in
Fig.~\ref{Fig:B-photon} is just the crossing diagram of the annihilation cross section).
%%%%%%%%%%%%%%%%%%%%%%%%%%%%
\begin{figure}[t]
\begin{center}
\subfloat{
\rotatebox{90}{\hspace{0.5cm} {Event rate/kgr/yr}}
% for $\sigma_N=10^{-6}$ pb}}
%\includegraphics[clip,width=1.0\linewidth]{Tot_127_0.eps}
\includegraphics[scale=0.6]{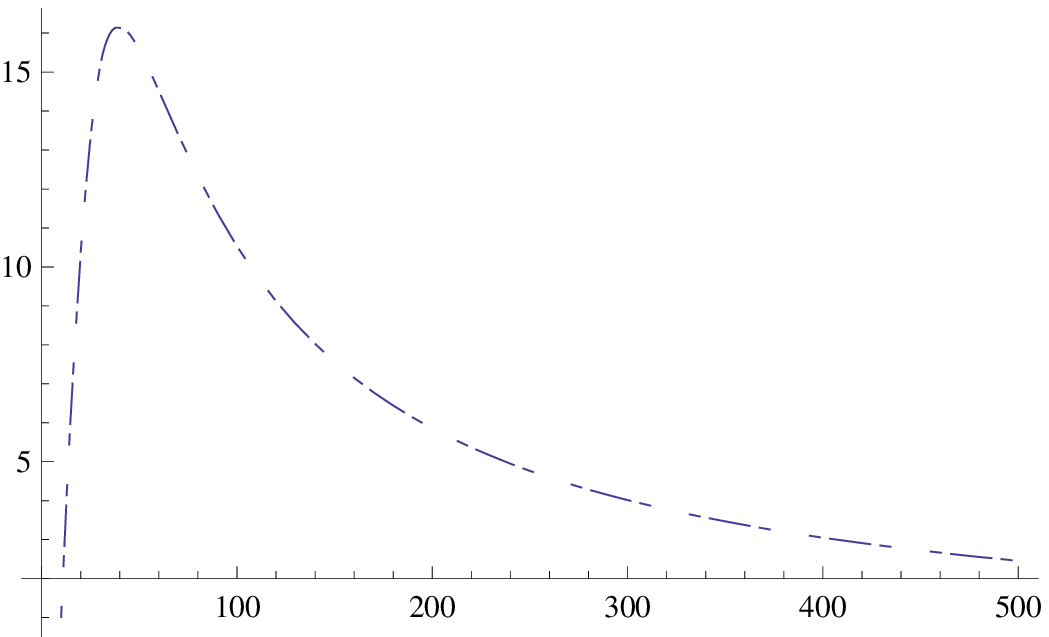}
}
\subfloat{
%{\hspace{5.0cm} $m_{\chi}\longrightarrow$  GeV}\\
\rotatebox{90}{\hspace{0.5cm} {Event rate/kgr/yr}}
% for $\sigma_N=10^{-6}$ pb}}
\includegraphics[scale=0.6]{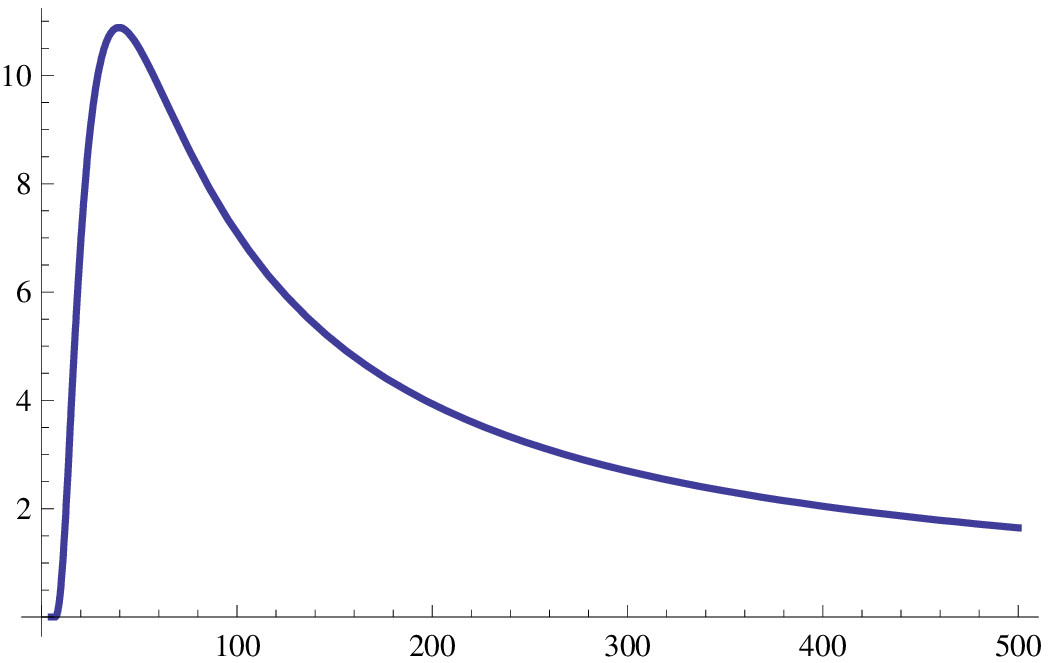}
}
{\hspace{5.0cm} $m_{\chi}\longrightarrow$  GeV}\\
{\hspace{-0.0cm} (a) \hspace{9.0cm} (b)}
\caption{\sl 
The total rates  for traditional WIMP searches  
assuming  a nucleon cross
section   $\sigma_N=10^{-43}$ $\mathrm{cm}^{2}$ in (a).
The case of the photon  mediated process 
considered in this work is exhibited in
(b). Both refer to the case of a heavy 
target (A=131) and were computed assuming
an energy threshold of 5 KeV. 
The results for the Iodine target used by the DAMA
experiment are almost identical.}
 \label{fig:Totrate_131}
 \end{center}
  \end{figure}
\begin{figure}[t]
\begin{center}
\subfloat{
\rotatebox{90}{\hspace{0.5cm} {Event rate/kgr/yr}}
% for $\sigma_N=10^{-6}$ pb}}
%\includegraphics[clip,width=1.0\linewidth]{Tot_127_0.eps}
\includegraphics[scale=0.6]{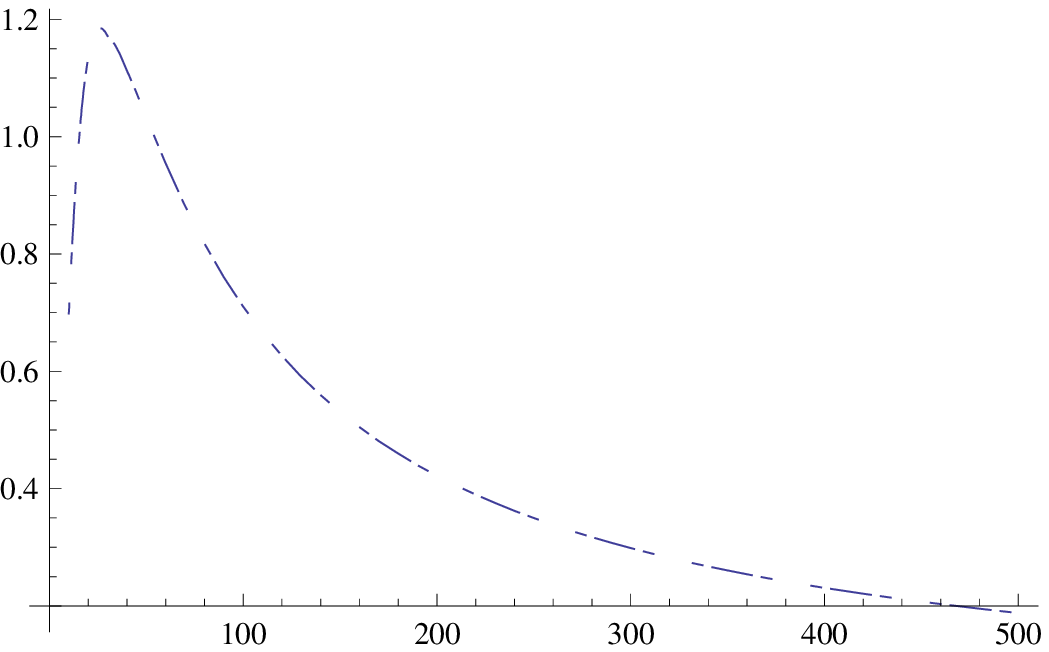}
}
\subfloat{
%{\hspace{5.0cm} $m_{\chi}\longrightarrow$  GeV}\\
\rotatebox{90}{\hspace{0.5cm} {Event rate/kgr/yr}}
% for $\sigma_N=10^{-6}$ pb}}
\includegraphics[scale=0.6]{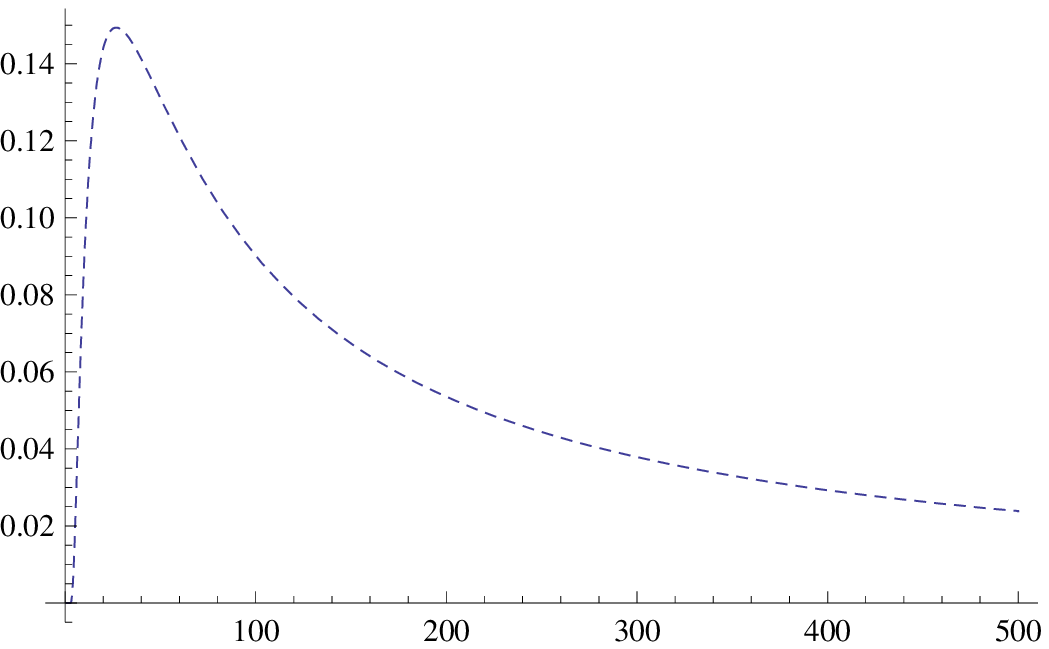}
}
{\hspace{5.0cm} $m_{\chi}\longrightarrow$  GeV}\\
{\hspace{-0.0cm} (a) \hspace{9.0cm} (b)}
\caption{\sl The same as in Fig. \ref{fig:Totrate_131} in the case of the light target $^{19}F$.}
 \label{fig:Totrate_19}
 \end{center}
  \end{figure}
%%%%%%%%%%%%%%%%%%%%%%%%%%

 Although \eq{bound1} [or \eq{bounddirac}] provides 
 a very stringent limit,  we should not forget that in 
 this case we have a much stronger dependence 
 of the rates  on the energy threshold through the 
 need for a low energy cut off on the elementary cross section.
  
\begin{figure}[t]
\begin{center}
\subfloat{
\rotatebox{90}{\hspace{0.0cm} {$ R(E_{th})/R(\left(E_{th} \right)_{min}\longrightarrow$ }}
% for $\sigma_N=10^{-6}$ pb}}
%\includegraphics[clip,width=1.0\linewidth]{th_stad_131.eps}
\includegraphics[scale=0.6]{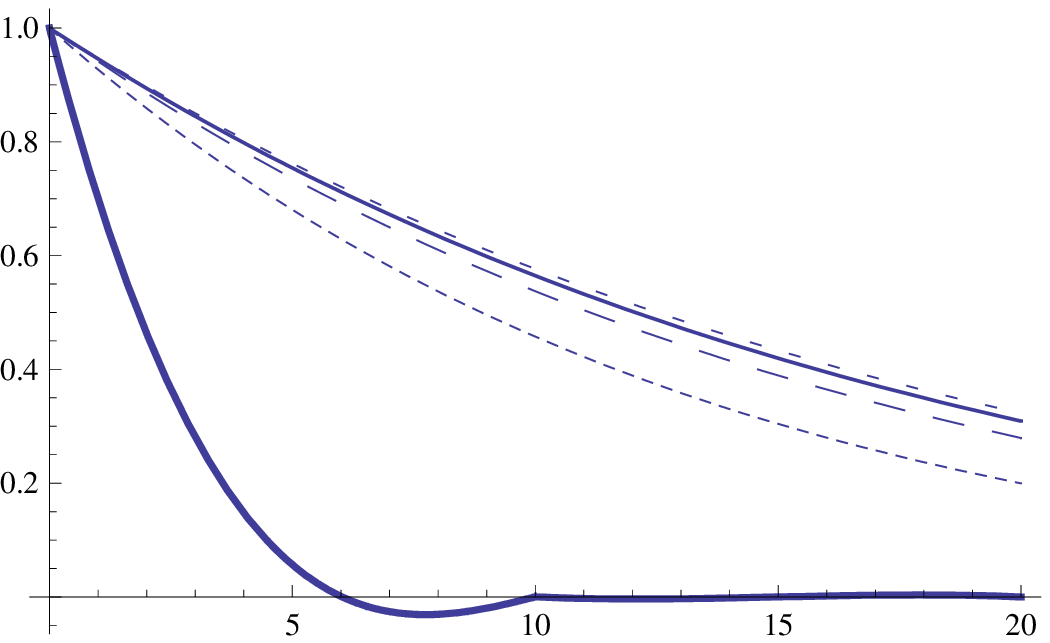}
}
\subfloat{
%{\hspace{5.0cm} $m_{\chi}\longrightarrow$  GeV}\\
\rotatebox{90}{\hspace{0.0cm}{$ R(E_{th})/R(\left(E_{th} \right)_{min}\longrightarrow$ }}
% for $\sigma_N=10^{-6}$ pb}}
\includegraphics[scale=0.6]{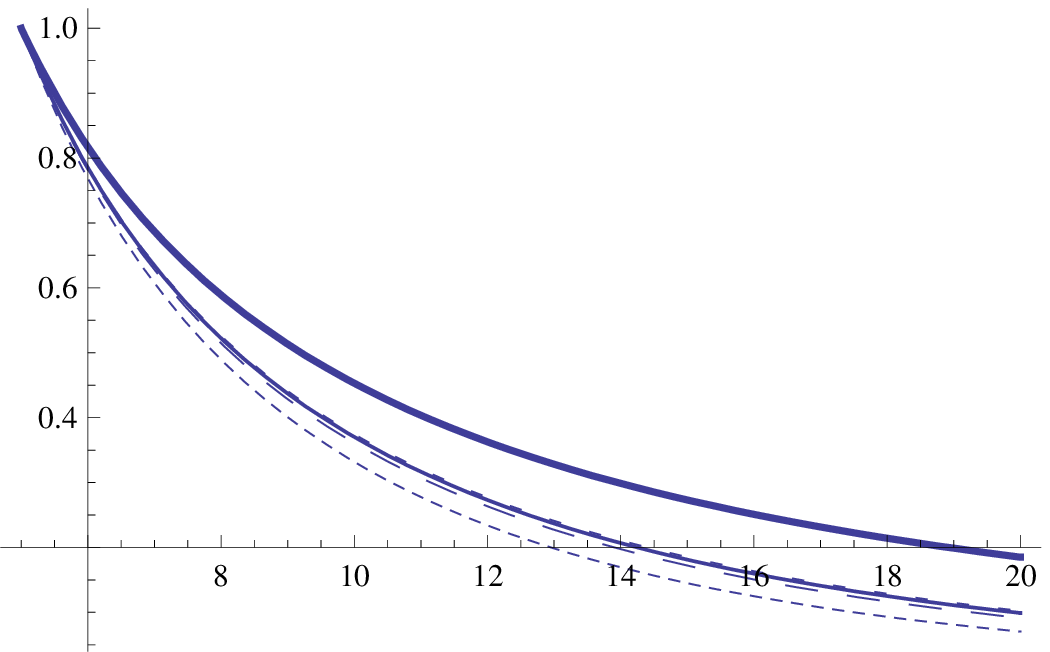}
}
{\hspace{6.0cm} $E_{th}\longrightarrow$  keV}\\
{\hspace{-0.0cm} (a) \hspace{9.0cm} (b)}
\caption{\sl The quantity
$R(E_\mathrm{th})/R(\left(E_\mathrm{th} \right)_{min}$, i.e. the ratio of the event rate at a  given threshold  divided by that at the lowest threshold considered, as a function of the threshold energy. In (a) as predicted by   traditional mechanisms (lowest threshold assumed zero). In (b) as predicted by  the present model (now due to the need for a cut off the lowest threshold energy employed was 5 keV). The thick line, short dash, long dash, fine line and long short dash correspond to WIMP masses 10, 50, 100, 200 and 500 GeV respectively. }
 \label{fig:th131}
 \end{center}
  \end{figure}

Alternatively  we may extract from the data for Xe  (A=131,Z=54) an elementary cross section assuming it to be of the form\footnote{This treatment does not distinguish between a Majorana and 
 a Dirac WIMP.}  
:
\beq
 \sigma_{N,\chi^0}^{S}\left(A,E_\mathrm{th} \right) \ = \
 \sigma_0 \frac{A}{131}\frac{5 ~\mathrm{keV}}{E_\mathrm{th}}\;,
\label{sigmanew}
\eeq
where $\sigma_0$ is the elementary cross section obtained in the particle model for a target with nuclear mass number $A$ and threshold energy $E_\mathrm{th}$. Then by fitting  to the experiment we obtain
\beq
(131/54)^2  \sigma_{N,\chi^0}^{S}=0.5\times 10^{-7}\Rightarrow \sigma_0=2.9\times 10^{-7}\mathrm{pb} = 2.9 \times 10^{-43}~\mathrm{cm}^{2}\;.
\eeq
In spite of the $\left(Z/A\right)^2$ factor we obtain a  smaller value than in the standard experiment. This is due to the small cut off energy $E_\mathrm{th}/A$ employed. 
%%%%%%%%%%%%%%%%%%%%%%%
 With the above ingredients the number of events in time $T$ due to the coherent scattering~\cite{TETRVER06}, can be cast in the form:
 %%%%%%%%%%%%%%%%%%%%
\barr
 R&&\simeq  1.07~10^{-5}\times
 \nonumber\\
 &&
\frac{T}{1 \mbox{y}} \frac{\rho(0)}{ \mbox {0.2~GeVcm}^{-3}}
\: \frac{100~\mbox{GeV}}{m_{\chi^0}}
\: \frac{m}{\mbox{1~kg}}
\: \frac{ \sqrt{\langle
%v^2 \rangle }}{280 {\mbox{km s}}$^{-1}$} \frac{\sigma_{p,\chi^0}^{S}}{10^{-6} \mbox{ pb}} f_{coh}(A, \mu_r(A))
v^2 \rangle }}{280~{\mbox {km s}}^{-1}}
\: \frac{\sigma_{N,\chi}^{S}}{10^{-43} \mbox{ cm}^{2}} \:
f_\mathrm{coh}(A, \mu_r(A))\;,
\label{eventrate}
\earr
%%%%%%%%%%%%%%%%%%%%
where the elementary cross section $ \sigma_{N,\chi}^{S}$can be treated as a phenomenological parameter independent of the WIMP mass in units of $10^{-43}$ $\mathrm{cm}^{2}$. The quantity $ f_\mathrm{coh}(A, \mu_r(A))$ can be obtained from the published  in \Rref{TETRVER06} values of $t$ for the standard MB velocity distribution (n=1).
For the  photon mediated  mechanism examined here the above equation must be modified by multiplying  $ f_\mathrm{coh}(A, \mu_r(A))$ with the factor  $Z^2/A^2$ and employing \eq{sigmanew} for the elementary cross section (in units of $10^{-43}$ $\mathrm{cm}^{2}$).
The event rate per kg of target per year for the traditional experiments for a heavy isotope like Xe and a light isotope like $^{19}$F, as a function of the WIMP mass is exhibited in Figs \ref {fig:Totrate_131} and \ref{fig:Totrate_19}. On the same plots we show the event rate for the photon mediated process examined in the present work. It is not surprising that the agreement is good since the elementary cross section was fitted to the data. The small difference is understood, since in the extraction of the elementary cross section from the data a zero threshold value was used in the phase space integrals.
% NEW TEXT
The event rates are sensitive functions of the threshold energy,
$R=R(E_\mathrm{th})$.  In the case of the Xe isotope
the ratio $ R(E_\mathrm{th})/R(\left(E_\mathrm{th} \right)_\mathrm{min}$
is exhibited in Fig. \ref{fig:th131}. The threshold dependence is much
more profound in the case of the light WIMP, since, then,
 the average energy transfered is small. As expected
 the threshold dependence is
more dramatic in the case of the present model (this is a bit
obscured in the figure since in this case the graphs are
normalized  at 5 keV).
%%%%%%%%%%%%%%%%%%%%%%%%%%%%

In the case of a Dirac fermion the extracted limit will be smaller, but the traditional calculations are not adequate for the analysis, due to the different velocity dependence of the elementary cross section.
% The present data seem to exclude such a possibility.

%%%%%%%%%%%%%%%%%%%%%%%%%%%%%%%%%%%%
\subsection{Massive Mediator}
\label{mm2}
%%%%%%%%%%%%%%%%%%%%%%%%%%%%%%%%%%%%

In this case the WIMP - nucleon cross section reads :
%%%%%%%%%%%%%%%%%%%%%%%
\begin{eqnarray}
\sigma \ &=& \ s(\beta) \, \frac{16\pi \aem \: \kappa^{2}\: \adm \:
m_{p}^{2}}{m_{X}^{4}}
\nonumber \\[3mm]
& = &
1.2 \times 10^{-30} \mbox{ cm }^{2}\, s(\beta) \: \frac{\alpha}{137^{-1}}\:
 \frac{\adm}{137^{-1}} \: \kappa^2 \, \left (\frac{m_p}{m_X} \right )^4 \;,
\end{eqnarray}
%%%%%%%%%%%%%%%%%%%%%%%%
where  the cross section refers to Dirac (Majorana) WIMP and
$s(\beta)=1(\beta^{2})$ respectively. Taking $\beta^2 \ \rightarrow \
< \beta^2 > \ \approx \ 10^{-3}$ we find:
\beq
\kappa\lesssim 3 \times 10^{-7}~(3 \times 10^{-4})\:.
\label{nc2}
\eeq
From these we obtain bounds for parameters in models I,II
 [see \eqs{gb}{eff2}] ,
%%%%%%%%%%%%%%%%%%%%%%%
\begin{eqnarray}
\epsilon & \lesssim & 3.0 \times10^{-7} \, (3.0 \times 10^{-3}) \;, \qquad \mathrm{Model~I}
\label{bo1}
\\[3mm]
Q_{X}\:
\frac{m_{Y}}{m_{X}} & \lesssim & 1.6 \times 10^{-6}\, (1.6\times 10^{-3})
\;, \qquad \mathrm{Model~II} \;,\label{bo2}
\end{eqnarray}
%%%%%%%%%%%%%%%%%%%%%%
where the number in parenthesis corresponds to Majorana WIMP
dark matter particle. These limits are less stringent than those obtained in the case of the massless mediator.\\
In the case of the massive mediator, with the possible exception of the velocity dependence in the case of Majorana WIMP, the cross section behaves as in the standard CDM case, since in this  case we do not encounter an energy cutoff.
 Since, however, we do not know the values of the parameters $\epsilon$ and $\frac{m_{Y}}{m_{X}} $, we cannot make predictions about the event rates. Instead we have used the present experimental limits to constrain these parameters. Thus we  saw that the current experimental limits impose the most stringent limits on these parameters. If, on the other hand, we use the previous constrains we can  conclude that WIMPs in models I,II
scatter off nuclei too many times. These
effects  should have been seen in experiments~\cite{CDMS,XENON}
 (or may have already been seen~\cite{DAMA}).
An exception is a Majorana WIMP candidate in model I which results
in current sensitivity event rates.

\section{Unconventional WIMP searches}
\label{sec:unco}

\subsection{Cross Section}

The other possibility is the direct scattering of WIMPs by electrons
that are bound in atoms.
The relevant Feynman diagram is depicted in Fig.~\ref{Fig:B-photon}
with quarks replaced by electrons. In this case only the electron flavor can be detected since the other flavors are not energetically allowed. Since the outgoing electrons are expected to  have energies in the eV region one cannot ignore atomic binding effects. The binding energy $b$ is found from the tables of ionization potential (energy) of an atom.\footnote{Tables are normally given in kJ/mol, but they can easily be translated in eV, since 96.485 kJ/mol = 1 eV. Thus for Cs we find $b=375.7/96.485=3.89$ eV.} 
%Sometimes we
%will find it convenient to express the binding energy
% in units of the electron's rest energy, $\tilde{b}=b/(m_ec^2)$.

The problem is to find the cross section for WIMP scattered off
an electron bounded in an atom.
In order to proceed we shall make two simplifying
assumptions :

\begin{enumerate}

\item As a working example, we shall assume
that the target is a hydrogenic atom denoted by $H$
 {\it i.e.,} a nucleus with
charge $+Ze$ and a single bounded electron with charge $-e$.
We shall discuss deviations from this assumption throughout.

\item The gauge boson mediator X couples
{\sl only} to WIMP and leptons but not to quarks. 
This is a necessary condition to
explain PAMELA positron excess of events. Therefore,
this discussion refers strictly to model III in \eq{Jxmu} 
[see however footnote 2].

\end{enumerate}

There are four processes that could take place in WIMP + H-like atom
collisions :
%%%%%%%%%%%%%%%%%%%
\begin{eqnarray}
\chi \ + \ H & \longrightarrow & \chi \ + \ H  \quad \mathrm{(elastic)}\;, 
\label{eq:0.1}\\
\chi \ + \ H  &\longrightarrow & \chi \ + \ H^{*} \quad \mathrm{(inelastic)}\;,
\label{eq:0.2} \\
\chi \ + \ H &\longrightarrow & \chi \ + \ e^{-} \ + \ H^{+}
\quad \mathrm{(production)}\label{eq0.3} \;.
\end{eqnarray}
%%%%%%%%%%%%%%%%%%%%%
For the rest we shall consider only the situation (\ref{eq0.3}).
The elastic scattering (\ref{eq:0.1}) cannot be detected,
and although we cannot exclude the inelastic one (\ref{eq:0.2}) 
from being experimentally probed through final state photons,
we believe that it would be
easier to detect the electrons from (\ref{eq0.3}). 
We shall assume that the electron emerges
with high momenta, ${\bf p_{e}'}$, such that in the final state its interaction
with the Coulomb potential in H-like atom is negligible, i.e, we can use
plane wave states for incoming and outgoing particles.
Using standard textbook~\cite{Peskin} wavepacket analysis  our
starting point will be the cross section formula in the lab frame:
%%%%%%%%%%%%%%%%%%%%%%%%%
\begin{eqnarray}
d\sigma \ &=& \
\frac{1}{2 E_{\chi} 2 E_{e}} \: \frac{1}{|v|} \:
\frac{d^{3} {\bf p_{\chi}'}}{(2\pi)^{3} 2 E_{\chi}'} \:
\frac{d^{3} {\bf p_{e}'}}{(2\pi)^{3} 2 E_{e}'} \:
|\overline{\mathcal{M}}|^{2} \: (2\pi)\: \delta(T_{\chi} - T_{\chi}' - T_{e}' -b)
\nonumber \\[3mm]
& \times &
d^{3}{\bf p_{e}} \: (2\pi)^{3}\:  \delta^{(3)}({\bf p_{\chi} + p_{e} - p_{\chi}' - p_{e}' }) \: |\phi(Z,{\bf p_{e}})|^{2} \;,
\end{eqnarray}
%%%%%%%%%%%%%%%%%%
where ${\bf p_{\chi}, p_{e}} ~({\bf p_{\chi}', p_{e}'})$
 are the incoming (outgoing)
three vector momenta of the WIMP and electron particles
respectively, and  $\overline{\mathcal{M}}$ is the matrix element
of the process $\chi + e \rightarrow \chi + e$ averaged over the
spins of the initial states  calculated in Born approximation.
We also ignore local velocity effects from the bound electron in the
(static in lab frame) atom {i.e., }
 that is the relative velocity  is $v \simeq v_{\chi}$.
$T_{i}=p_{i}^{2}/2m_{i}, i=\chi, e$ are the kinetic energies and $b$
is the binding energy of the electron in H-atom ($\approx 13.6$ eV).
Moreover, in non-relativistic limit
$E_{\chi} \simeq E_{\chi}' \approx m_{\chi}$ and $E_{e} \simeq E_{e}' \approx m_{e}$ with $m_{\chi} \gg m_{e}$,
while $\phi_{n\ell m_{\ell}}({\bf p})$, normalized at $\int_{V}
d^{3} p |\phi_{n\ell m_{\ell}}({\bf p})|^{2}=1$, is the Fourier transform
of the coordinate wave function $\psi_{n\ell m_{\ell}}({\bf r})$.
Using the $\delta^{(3)}$-function to perform the integration over ${\bf p_{e}}$, we obtain:
%
%%%%%%%%%%%%%%%%
\begin{eqnarray}
d\sigma \ = \
\frac{|\overline{\mathcal{M}}|^{2}}{16 m_{\chi}^{2} m_{e}^{2} \beta} \:
\frac{d^{3}{\bf p'_{\chi}} \: d^{3} {\bf p_{e}'}}{(2\pi)^{2}}\,
\delta \biggl(\frac{|{\bf p_{\chi}}|^{2}}{2m_{\chi}}
 - \frac{|{\bf p_{\chi}'}|^{2}}{2 m_{\chi}} -
\frac{|{\bf p_{e}'}|^{2}}{2 m_{e}} - b(Z)\biggr  )\,
|\phi_{n\ell m_{\ell}}(Z,{\bf p_{\chi}' + p_{e}' -p_{\chi}})|^{2} \;,
\label{eq4.37}
\end{eqnarray}
%%%%%%%%%%%%%%%%%%
where the energy conservation delta-function has been
written out explicitly.
The result of \eq{eq4.37} is a product of two parts : a part that
 contains the dynamics of the WIMP-electron interaction
 through the matrix element $|\overline{\mathcal{M}}|$
times the probability of finding the target electron with momentum
${\bf p_{e}} = {\bf p_{\chi}' + p_{e}' -p_{\chi}}$ in H-atom.
In addition the matrix element
of the process $\chi + e \rightarrow \chi + e$ averaged over the
spins of the initial states in Born approximation
reads :
%%%%%%%%%%%
\begin{eqnarray}
|\overline{\mathcal{M}}|^{2} \ \simeq \
\frac{(16 \pi)^{2} \adm \alpha' m_{e}^{2}
m_{\chi}^{2}}{({\bf |p_{\chi}-p_{\chi}'}|^{2}-m_{X}^{2})^{2}} \: s(\beta) \;,
\end{eqnarray}
%%%%%%%%%%%
where the factor
$s(\beta) \equiv 1 \: (\beta^{2})$ for
Dirac WIMP (Majorana WIMP) particle.
Note that the cross section for Majorana WIMP is always smaller
by a factor of $\beta^{-2}$ compared to the one involving Dirac WIMP.
We now use the kinetic energy $\delta$-function appearing in \eq{eq4.37}
in order to perform the $|{\bf p_{\chi}'}|$ integration and
arrive at:
%
%%%%%%%%%%%
\begin{eqnarray}
d\sigma \ = \ s(\beta) \: \frac{16 \pi^{2} \adm \alpha'
m_{\chi}^{2}}{({\bf |p_{\chi}-p_{\chi}'}|^{2}-m_{X}^{2})^{2}} \:
\frac{|{\bf p_{\chi}'}|}{|{\bf p_{\chi}}|} \:
|{\bf p_{e}'}|^{2} d |{\bf p_{e}'}| \: |\phi_{n\ell m_{\ell}}(Z,{\bf p_{\chi}'+p_{e}'}-{\bf p_{\chi}})|^{2}\:
d \xi \: d\eta \;,\label{4.39}
\end{eqnarray}
%%%%%%%%%%%%%%%%%%
where the initial WIMP momentum is
$|{\bf p_{\chi}}| = m_{\chi} \beta$ and
 the scattering angles are defined as
%%%%%%%%%%%%%%%%
\begin{eqnarray}
\xi = \hat{p}_{\chi} \cdot \hat{p}_{\chi}'  \;, \qquad
\eta = \hat{p}_{\chi} \cdot \hat{p}_{e}'  \;, \qquad \xi ,\eta \in [-1,1]\;.
\end{eqnarray}
%%%%%%%%%%%%%%%%
The integration over the azimuthal angles has been carried out trivially
in \eq{4.39} and
the momentum $|{\bf p_{\chi}'}|$ of the scattered WIMP 
is found to be
%%%%%%%%%%%%%%%%
\begin{eqnarray}
|{\bf p_{\chi}'}| \ = \ \sqrt{m_{\chi}^{2}\: \beta^{2} \ -\ 2 m_{\chi}\: b(Z)
\ -\ \frac{m_{\chi}}{m_{e}}\: p_{e}^{'2}} \;, \quad \mathrm{with}
 \qquad p_{e}' \ = \
\sqrt{2 m_{e} E_{e}'} \;, \label{eq4.46}
\end{eqnarray}
%%%%%%%%%%%%%%%%%%%%%%%
where $b(Z)$ is the ground state energy for hydrogenic atoms is
%%%%%%%%%%%%%%%%
\begin{eqnarray}
b(Z) \ = \ \frac{Z^{2}}{2 a}\frac{e^{2}}{4\pi} \ =\
\frac{Z^{2}}{2} m_{e} \: \aem^{2}
 \;, \qquad a \simeq \frac{1}{m_{e} \: \aem} \;, \label{4.43}
\end{eqnarray}
%%%%%%%%%%%%%%%%%%%%%%%
in the
approximation  $\mu \simeq m_{e}$ where $\mu$ is
the reduced mass,
with $\aem = \frac{e^{2}}{4\pi} \approx 1/137$, $m_{e} \simeq 0.5$ MeV
and $a = a_{0} \approx 0.5\: \mbox{\AA}$ being
the Bohr radius for $Z=1$.
Throughout this chapter, we are going to use the ground
state momentum distribution  of
hydrogenic atoms which reads:
%%%%%%%%%%%%%%%%
\begin{eqnarray}
\phi_{100}(Z,p) \ = \ \frac{2^{3/2}}{\pi a} \: \frac{(Za)^{5/2}}{(Z^{2} + p^{2}a^{2})^{2}} \;.
\end{eqnarray}
%%%%%%%%%%%%%
Notice that since $\phi_{100}(p)$ depends on $|p|^{2}$
and therefore from the scattering angles $\eta$ and $\xi$
and electron energy $E_{e}'$.
A term in \eq{4.39},
$\frac{|{\bf p_{\chi}'}|}{|{\bf p_{\chi}}|} =
\frac{|{\bf v_{\chi}'}|}{|{\bf v_{\chi}}|} $,
arises from the fact that we treated the H-atom as a brick wall
potential. Had we not done so, the influence of the Coulomb
potential on the emerging electron would not have been uniquely
correlated to ${\bf p_{\chi}'}, {\bf p_{\chi}}$
and the back reaction of the proton
should have been taken into account.

Exactly the same result as in \eq{4.39} can be found by using simpler
time-dependent perturbation  theory for transitions to continuum
in non-relativistic quantum mechanics~\cite{Schiff}.
In a more refined analysis however, when 
the recoiling energy is in the neighborhood of 
the binding energy of the atom
one should take into account effects from the continuum hydrogenic
wave functions instead of treating the final electron as 
plane wave. This analysis, though more accurate, is far more complicated and
does not change the qualitative features of our results.
It can be addressed in the future (together with  other effects)
especially if these kind of experiments become operative [see subsection
4.4 below].

We
analyze below the corresponding cross sections for
a  massless and a massive mediator as we did
 in section~\ref{sec:con} for the nucleons.

%%%%%%%%%%%%%%%%%%%%%%%%%%%%%%%%%%%%
\subsubsection{Event Detection Rates}
\label{sec:er}
%%%%%%%%%%%%%%%%%%%%%%%%%%%%%%%%%%%

In general for an atom, due to binding energy effects only the loosely bound electrons can contribute to the process (\ref{eq0.3}). So we will convolute the elementary cross section with the WIMP velocity distribution, which, with respect to the galactic center, we will take to be Maxwell-Boltzmann form:
%%%%%%%%%%%%%%%%%%%%%%%%%%%%
\beq
f(\beta) \ = \
\left (\frac{3}{2<\beta^2>} \right )^{3/2}\:
\frac{1}{ \pi ^{3/2}} \: e^{-\frac{3 \beta ^2}{2
   <\beta^2>}}\;.
\eeq
%%%%%%%%%%%%%%%%%%%
Transforming this into the local coordinate system:
%%%%%%%%%%%%%%%%%%%%%
\begin{equation}
\beta\rightarrow\beta \hat{\beta}+\beta_0\hat{z}=
\beta\hat{\beta}+\sqrt{\frac{2<\beta^2>}{3}}\hat{z} \quad , \quad
\beta^2\rightarrow \beta^2+\frac{2}{3}<\beta^2>+
2 \beta \cos(\theta) \sqrt{\frac{2}{3}<\beta^2>} \;,
\end{equation}
%%%%%%%%%%%%%%%%%%%
where $\theta$ is the  angle between $\hat{\beta}$ and $\hat{z}$
and
 $\beta_0=\sqrt{\frac{2<\beta^2>}{3}}$ is the sun's velocity with respect to the center of the galaxy and $<\beta^2>\approx 10^{-6}$.
Then we obtain the local
distribution of speeds $f_{\ell}(\beta)$ relative to the detector to be
%%%%%%%%%%%%%%%%%%%%
\beq
f_{\ell}(\beta)=\left (\frac{3}{2<\beta^2>} \right )^{3/2}\frac{1}{ \pi ^{3/2}}
e^{-\left (\frac{3 \beta ^2}{2
   <\beta^2>} \ + \ 2\, \beta\, \cos(\theta)\, \sqrt{\frac{3 }{2
   <\beta^2>}} \ + \ 1 \right )}\;.
\eeq
%%%%%%%%%%%%%%%%%%%%%%%
The integration over the angles of the distribution can be done analytically. In evaluating the rate one has to incorporate the oncoming flux.
So, adopting  appropriate  normalization,
in the  convolution we introduce the factor $1/\sqrt{<\beta^2>}$.
This way we find the rate to be proportional  to :
%%%%%%%%%%%%%%%%%%%%%%%%%%%%%
\beq
\frac{\beta\, f_{\ell}(\beta) \, d\beta}{\sqrt{<\beta^{2}>}} \ = \
\left (\frac{3}{2<\beta^2>} \right )^{3/2}\:
\frac{2}{ \sqrt{\pi}} e^{-\left (\frac{3 \beta ^2}{2
   <\beta^2>} +1 \right )} \frac{\beta^3}{\sqrt{<\beta^2>}}
   \frac{\sinh\left(2 \beta \sqrt{3/(2<\beta^2>})\right )}{ \beta
   \sqrt{3/(2<\beta^2>)}}\: d\beta \;. \label{eq4.47}
\eeq
%%%%%%%%%%%%%%%%%%

Combining this with the cross section of \eq{4.39}
obtained previously we arrive at:
%%%%%%%%%%%%%%%%%%%%%%%%%%%%
\beq
\left \langle \frac{d \sigma}{dE_{e}'}
 \frac{\beta}{\sqrt{<\beta^{2}>}} \right \rangle
\ = \ \int_{\beta_{\rm min}}^{\beta_{\rm esc}} d\beta\: \frac{\beta
f_{\ell}(\beta)}{\sqrt{<\beta^{2}>}} \: \frac{d \sigma}{dE_{e}'} \;,
\eeq
%%%%%%%%%%%%%%%%%%%%%%
where the lower velocity in the integral can be read from the
positivity of the square root quantity in \eq{eq4.46}
%%%%%%%%%%%%%%%%%%%%%%%
\beq
\beta_{\rm min} = \sqrt{\frac{2 E_{e}'}{m_{\chi}} +
\frac{2 b(Z)}{m_{\chi}}}\;,
\eeq
%%%%%%%%%%%%%%%%%%%%%%%%%%%%%%
and $\beta_{{\mathrm esc}}=2.84\sqrt{(2/3)<\beta^2>} $ is the escape velocity.
It is now easy to calculate the differential event rate per eV
ejected electron energy per year and per kilogram of target
material to be
%%%%%%%%%%%%%%%%%%%%%%
\beq
\frac{dR}{dE_{e}'} \ = \ \frac{\rho_0}{m_{\chi}} \sqrt{<\beta^2>} N_e \left \langle \frac{d\sigma}{dE_{e}'}\:
\frac{\beta}{\sqrt{<\beta^{2}>}} \,\right  \rangle \;,\label{4.50}
\eeq
%%%%%%%%%%%%%%%%%%%
where $\rho_{0}=0.2~ \mathrm{GeV}/\mathrm{cm}^{3}$ is the
WIMP energy density and $N_{e}$ is the number of target electrons.
Integration of \eq{4.50} upon $E_{e}'$ over the region
from $E_{e_{min}}'=0$ to $ [m_{\chi} \beta_{\rm esc}^{2}/2-b(Z) ]$
 results in the total event number per unit time
and mass of the target which among other parameters depends
on the mass and atomic numbers of the  target atom.
Moreover we shall display results 
 on the total event rate $R(Z)$ when $E_{e_{min}}' = E_{\rm th}$
 with varying experimental threshold energy $E_{\rm th}$.

\subsubsection{Time Modulation Effects for Electrons}

In the convolution of the elementary cross section we have so far
considered only the motion of the sun with respect to the center
of galaxy. More realistically, one should consider also the Earth's
velocity and then find the modulated event rate that might be detected
on  Earth. In this case the WIMP velocity is read from
%%%%%%%%
\begin{eqnarray}
{\bf v'} \ = \ {\bf v} \ + \ v_{0}\, {\bf {\hat{z}}}\ + \ v_{1}\, (\sin{\alpha}\: {\bf
{\hat{x}}} \ + \ \cos{\alpha}  \cos{\gamma}\: {\bf{\hat{y}}} \ + \ 
\cos{\alpha}  \sin{\gamma} \, {\bf{\hat{z}}})
 \;, \label{velocity}
\end{eqnarray}
%%%%%%%%%%%%%%%%%%%%%%%
where $v_{0}$ is Sun's velocity, $v_{1}$ is Earth's annual velocity,
$\gamma=\frac{\pi}{6}$ is the angle between the projection of
vector ${\bf{v_{1}}} $ on the plane $yOz$ and the ${\bf
{\hat{y}}}$ direction 
and $\alpha = a(t)$ is the complementary angle of the angle
between $\bf {v_{1}}$ and ${\bf {\hat{x}}}$. Then the WIMP cross
section has to be convoluted with 
%%%%%%%%%%%%%%%%%%%%%%%%%%%%%
%\beq
%\frac{\beta\, f_{\ell}(\beta) \, d\beta}{\sqrt{<\beta^{2}>}}
%\ = \ \left (\frac{3}{2<\beta^2>} \right )^{3/2}\: \frac{2}{
%\sqrt{\pi}}\; e^{-\left (\frac{3 \beta ^2}{2
%  <\beta^2>} +1 \right )} \beta^2\sqrt{\frac{2}{3}}
%   \sinh\left(2 \beta \sqrt{\frac{3}{2<\beta^2>}}\right )\left(1+ k
%  \delta \cos{\alpha}\right) d\beta
%  \;,
%\eeq
%%%%%%%%%%%%%%%%%%
\beq
\biggl (\frac{\beta\, f_{\ell}(\beta) \, d\beta}{\sqrt{<\beta^{2}>}} 
\biggr ) \ = \ \biggl (\frac{\beta\, f_{\ell}(\beta) \, d\beta}{\sqrt{<\beta^{2}>}} 
\biggr )_{0} \: (1 + k\, \delta \, \cos \alpha) \;,
\eeq
%%%%%%%%%%%%%%%%
where the expression with the subscript ``0'' refers to \eq{eq4.47}
with  $\delta=\frac{v_{1}}{v_{0}}\approx0.135$  and
%%%%%%%%%%%%%%%%%%%%%%%
\beq
k \ = \ \left(2\beta \sqrt{\frac{3}{2< \beta^{2}> }}\;
 \frac{\cosh\left({2\beta
\sqrt{\frac{3}{2<\beta^{2}>}}}\right)
}{\sinh{\left(2\beta\sqrt{\frac{3}{2<\beta^{2}>}}\right)}}
-3\right)\sin{\gamma} \;. \label{4.53}
\eeq
%%%%%%%%%%%%
It is now trivial to extend the distribution with energies event rate of
\eq{4.50} with
 %%%%%%%%%%%%%%%%%%%%%%%%%%%%
 \beq
 \frac{dR}{dE_{e}'} \ = \ \left<\frac{dR}{dE_{e}'}\right>_{0} \ + \
 \left<\frac{dR}{dE_{e}'}\right>_{\rm mod} \times  \cos{\alpha}
   \; \label{tmmassless}
\eeq 
%%%%%%%%%%%%%%%%%%%%
where $\left<\frac{dR}{dE_{e}'}\right>_{0}$ is the
unmodulated differential event rate while $\left<\frac{dR}{dE_{e}'}\right>_{\rm mod}$ contains also the factor $k$ in \eq{4.53}.
%%%%%%

\subsection{Massless Mediator}
\label{sec:mm3}

In this case
dark matter scattering happens
 via the coupling of the exotic gauge boson to the photon (model II).
 The Feynman diagram is identical to the one presented in Fig.
\ref{Fig:B-photon} with the quark replaced by electron.
In general case  the WIMP-electron cross section is not independent of the velocity. Thus, we will first estimate the cross section by using an average velocity $\sqrt{<\beta^2>}=10^{-3}$.
 Following
\eq{4.39} for a photonic mediator we find the
differential cross section:
%%%%%%%%
\begin{eqnarray}
\frac{d\sigma}{dE_{e}'}\ = \  s(\beta) 16 \pi^{2}\alpha' \adm
\:\kappa^2\: m_{\chi}^{2} m_{e} \: \frac{|{\bf p_{\chi}'}|}{|{\bf
p_{\chi}}|} \: |{\bf p_{e}'}| \: \int_{-1}^{1} d\xi \:
\int_{-1}^{1} d\eta \: \frac{|\phi_{n\ell m_{\ell}}(Z,{\bf
p_{\chi}'+p_{e}'-p_{\chi}})|^{2}}{({\bf p_{\chi}'}-{\bf
p_{\chi}})^4}
 \;, \label{eq4.55}
\end{eqnarray}
%%%%%%%%%%%%%%%%%%%%%%%
where ${\bf q} ={\bf p_{\chi}'}-{\bf p_{\chi}}$ is the WIMP
momentum transfer which is $\xi$ dependent. The cross section
peaks up the most from the forward direction $\xi \approx 1$. 
It should be mentioned that since the initial electron is bound,
there is no infrared divergence in this case.
Moreover, the 
momentum transfer can be as low as :
%%%%%%%%%%%%%%%
\beq
|{\bf q}| \ \simeq \ 2\:  \frac{ b(Z) + E_{e}'}{\beta} \;.
\label{momtra}
\eeq
%%%%%%%%%%%%%%% 
This relation is important for explaining our numerical results below.
Furthermore, in presenting the results we
assume a Dirac WIMP fermion,i.e. $s(\beta)=1$. Furthermore,
we choose a benchmark scenario inspired by our findings in nucleon 
decay :
\beq
\beta =\sqrt{<\beta^{2}>} = 10^{-3}\;, \quad Z=1 \;, \quad  \adm = \alpha' = \alpha_{\mathrm{em}} \;,  \quad
m_{\chi}=100~\mathrm{GeV} \;, \quad\kappa=10^{-10} \; .
\label{4.56} \eeq
As it is obvious from \eq{eq4.55} it is very easy to 
apply our numerical results to any other  
parameters, $\beta,\adm, \alpha', \kappa$,
 than those shown in \eq{4.56}. 
 We must note here that there is no parameter $\kappa$ in model III.
 This parameter is used here as a rescale factor  and its
 very small value is adjusted so that we  obtain rates
 of few events. 

\begin{figure}[t]
   \centering
   \hbox{\includegraphics[height=2in]{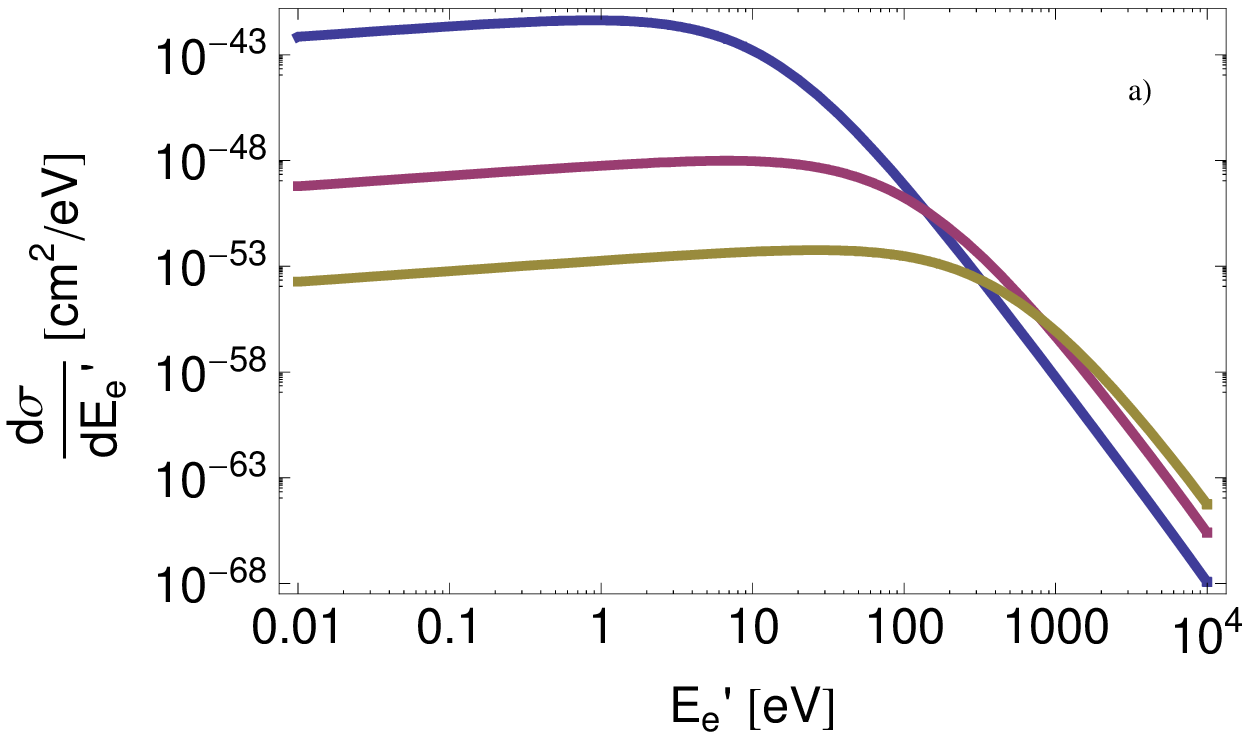}\hspace*{0.5cm}
    \includegraphics[height=2in]{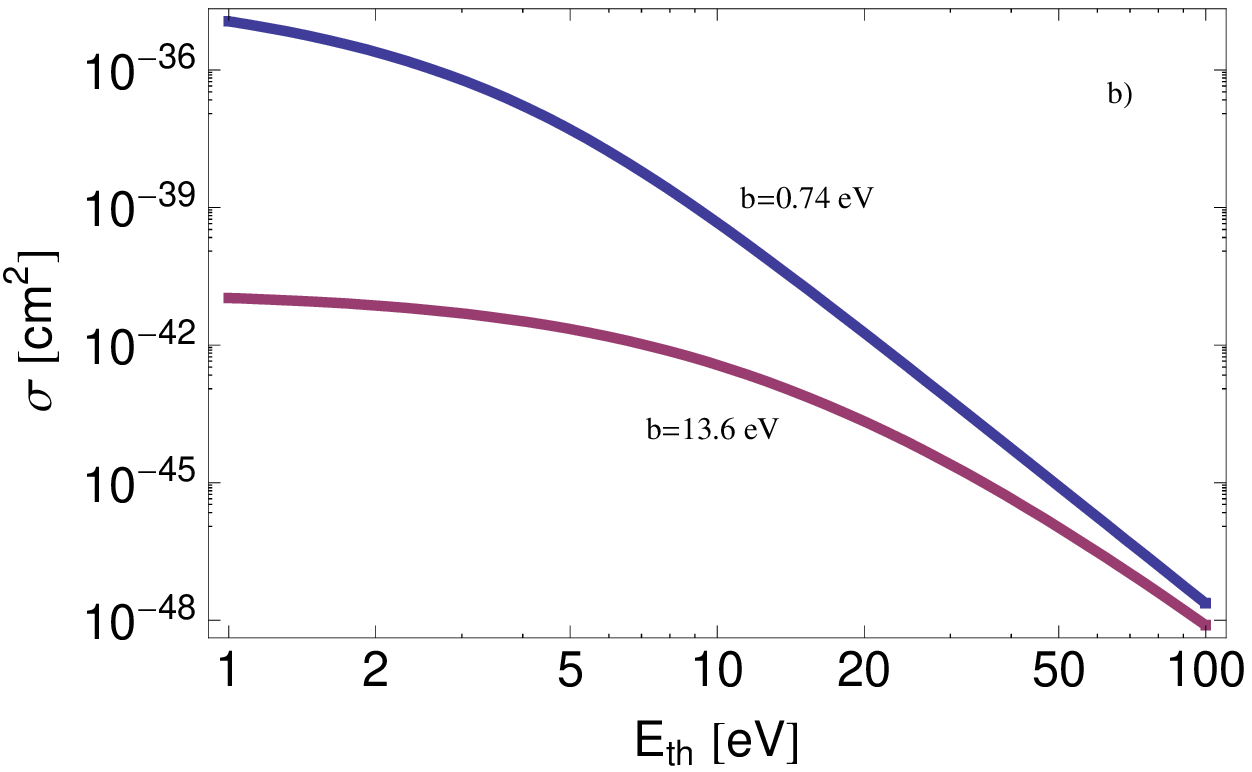}}
    \includegraphics[height=2in]{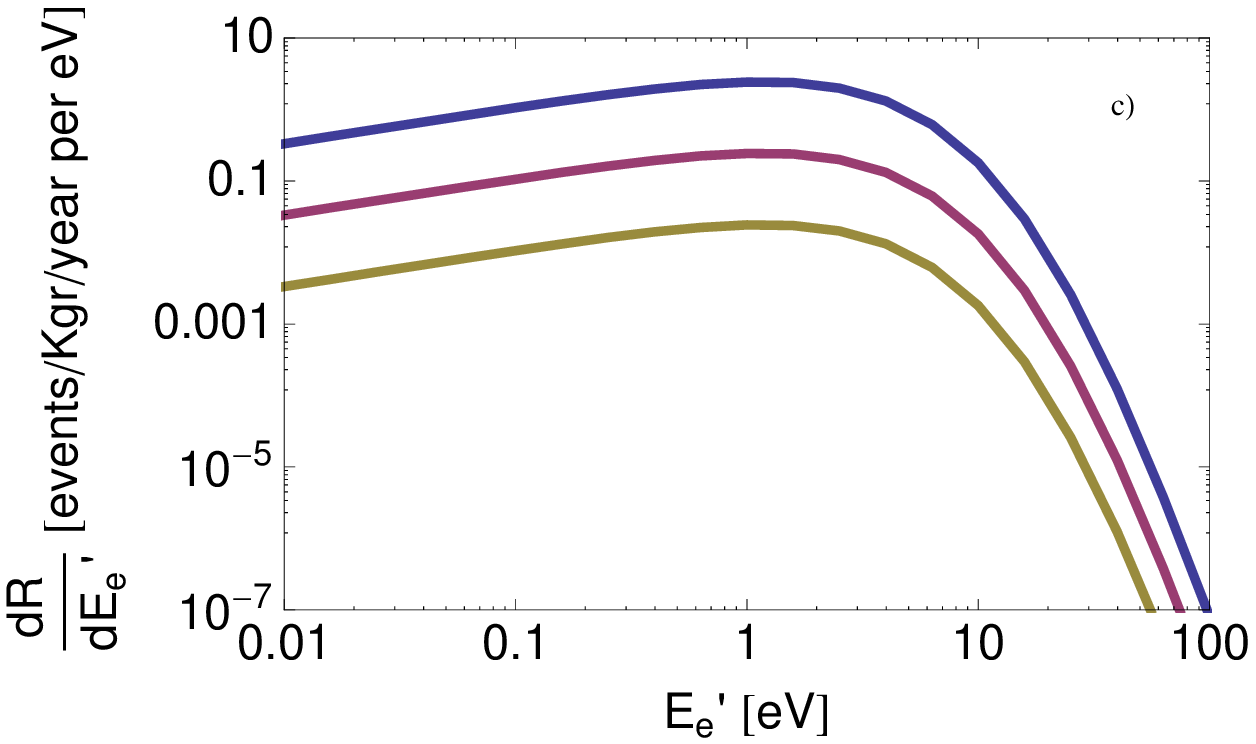} 
   \caption{\sl a) Predictions for $d\sigma/dE_{e}'$ as a function of
   the ejected electron energy $E_{e}'$. The  target is assumed to be a 
   hydrogenic atom in the ground state with  Z=1,3,6 (from top to bottom).
   b) The total cross section for process (\ref{eq0.3}) as a function of the     
   experimental threshold energy for two binding energies.
   c) The differential event rate as a function of the electron energy 
   and various WIMP masses (10,100,1000) GeV from (top to bottom). 
   Other parameters not shown, are taken  from \eq{4.56}.  }
   \label{fig:dsdemassless}
\end{figure}

 In Fig.~\ref{fig:dsdemassless}a are shown the results for
the $d\sigma/dE_{e}'$ as a function of final electron's energy
$E_{e}'$ for three different cases of hydrogenic atoms with $Z=1 \,,
Z=3$ and $Z=6$ respectively.
 %%%%%%%%%%%%%%%%
The differential cross section takes on its maximum values
 for final electron energy
 of around few eV for $Z=1$, around few tens  of eV for $Z=3$ and around
 a hundred  eV for $Z=6$. For the case $Z=1$, the extremum
  happens because of a fast increase of the
 term $\frac{|{\bf p_{\chi}'}|}{|{\bf p_{\chi}}|} \:
|{\bf p_{e}'}| \sim \sqrt{E_{e}'}$    and the almost constant
value of $|\phi_{100}|^{2}$ until 5 eV . For higher electron
energies, e.g., $E_{e}' \gtrsim 10$ eV,  the probability density
factor $|\phi_{100}|^{2}$  drops fast as $1/E^{'8}_{e}$ and the
term in the denominator of the integral increases as $E^{'2}_{e}$,
resulting in overall decreasing of the cross section as
$E_{e}^{'-19/2}$.The same analysis can be used to describe the
behavior of $d\sigma/dE_{e}'$ in the other cases $(Z=3,Z=6)$.
We must note here the in the limit $E_{e}' \to 0$ we obtain
$d\sigma / dE_{e}' \to 0$ as the case  should be. This is 
obscured in Fig.~\ref{fig:dsdemassless} due to the range 
choice of $E_{e}'$.
 %%%%%%%%%%%%%

%%%%%%%%%%%%%%%%%%%%%%%%%%%%%%%%%
%%%%%%%%%%%%%%%%%%%%%%%%%%%%%%%%%

Corresponding to the input parameters noted in (\ref{4.56}) we
calculate the total cross section from \eq{eq4.55} after numerical
integration over $E_{e}'$ in the region $[E_{\rm th}, m_{\chi} \beta^{2}/2
- b(Z)]$. Our results for $\sigma$ vs.  the threshold energy 
$E_{\rm th}$ are depicted in Fig.~\ref{fig:dsdemassless}b.
We have chosen two extreme cases of binding energies :
$b=0.74$ eV that is the the binding energy of the electron
bounded in the two electron atom $H^{-}$ 
and $b=13.6$ eV that is the one corresponding
to the H- atom we have been dealing so far. For 
$E_{\rm th}\lesssim 10$ eV the difference in cross section 
is about three to six  orders of magnitudes while for 
higher threshold energies becomes unimportant.

Following \eq{eq4.55} it turns out that the 
total cross section for process (\ref{eq0.3}) is WIMP mass independent.
 It is experimentally useful to know how  the cross section depends on
the threshold energy $E_\mathrm{th}$ 
that a given experiment can accomplish.
This  
is plotted in Fig.~\ref{fig:dsdemassless}b.
For $E_{\rm th} \lesssim 1$ eV, the cross
section is essentially independent of $E_{\rm th}$.
When the threshold becomes 5 eV, in the case of $b(Z)=13.6$ eV,
the cross section drops
by a factor of 5 eV while 
up to 10 eV by a factor of 50. For smaller binding energy though, 
i.e., $b(Z)=0.74$ eV,  and up to 10 eV the
cross section decreases  by three orders of magnitude.

Furthermore, the dependence of differential event rate
$dR/dE_{e}'$ as a function of  the ejected electron energy $E_{e}'$ for three
different WIMP masses, $m_\chi=10,100,1000$ GeV, is shown in 
 Fig.~\ref{fig:dsdemassless}c. There is a maximum
which follows the behavior of differential cross section. 
The event rate falls as $1/m_{\chi}$ as the WIMP mass increases
in accordance with \eq{4.50}. For energy of few eV's and $m_{\chi}= 10$ GeV we obtain a handful of events for $\kappa = 10^{-10}$.
A total event rate is obtained after integrating 
over the differential rate
in Fig.~\ref{fig:dsdemassless}c. 
As a typical value, for $m_{\chi}= 100$ GeV  and the
parameters in (\ref{4.56})
we find  $R(Z=1,\kappa=10^{-10}) \approx 1$ events/yr/target~kgr.
The reader must recall here  that this assumes
 a mixing parameter as small as $\kappa = 10^{-10}$ !!

Finally, following the theoretical discussion of the previous subsection 
we  examine effects of the WIMP time modulation. 
In  Table~1  we display
both the unmodulated and modulated differential event rate for four
representative values of  $E_{e}'$ in  the case of a massless
mediator and parameters of \eq{4.56}. 
The dimensionless parameter $H$, which is the ratio of the modulated by the non modulated differential amplitude, is constant around $9-13 \%$ independent of the energy and the WIMP mass. So the  modulation 
$h=\delta \cdot k $ of the total rate is also going to be around $10 \%$, which means that the difference between the maximum (here always in June 3rd) and the minimum (here  always in December) is 
$18-26 \%$, a result should
not to be overlooked.

%%%%%%%%%%
%%%%%%%%%%%%%%%%%%%%%
\begin{table}[t]
\label{table1}
\begin{center}
\begin{tabular}{cccc}
\hline
\multicolumn{1}{c}{$E_{e}' $ [eV] }&
\multicolumn{2}{c}{$\left<\frac{dR}{dE_{e}'}\right>$ [events/kgr\,
target/year/eV]}\\
  & unmod. & mod. & H
 \\\hline
$0.1$ & $0.11$  & $0.01$ & 0.09
\\ $1$ & $0.24$ & $0.03$ & 0.13 \\ $10$ & $0.02$ & $0.002$ & 0.10
\\ $100$ & $8.21\times 10^{-9}$ & $1.04\times 10^{-9}$ & 0.13
\\\hline
\end{tabular}
\caption{\sl Time modulation effects  in case of a photonic mediator
following \eq{tmmassless} in the text.
Various input parameters are given  in \eq{4.56}.  H is the ratio 
of the modulated divided by the unmodulated differential rate.}
\end{center}
\end{table}
 %%%%%%%%%%%%%%%%%%%%%%%%%%%%%%

\subsection{Massive mediator}
\label{sec:mm4}
%%%%%%%%%%%%%%%%%%%%%

Again the relevant diagram is the one of Fig.~\ref{Fig:B-photon}
with quarks replaced by electrons.
By  taking the non-relativistic limit of \eq{4.39} and the assumption that the
momentum transfer in \eq{momtra}
is much less than the mediator mass, $q^{2} \ll m_{X}^{2}$, we arrive at
%%%%%%%%%%%%%%%%
\begin{eqnarray}
\frac{d\sigma}{dE_{e}'}\ = \  s(\beta)
\frac{16 \pi^{2}\alpha' \adm \kappa^{2}}{m_{X}^{4}}\:
m_{\chi}^{2} m_{e} \:
\frac{|{\bf p_{\chi}'}|}{|{\bf p_{\chi}}|} \:
|{\bf p_{e}'}| \:  \int_{-1}^{1} d\xi \: \int_{-1}^{1} d\eta \:
|\phi_{n\ell m_{\ell}}(Z,{\bf p_{\chi}'+p_{e}'-p_{\chi}})|^{2}
 \;, \label{eq1.20}
\end{eqnarray}
%%%%%%%%%%%%%%%%%%%%%%%
In what follows we assume a Dirac WIMP fermion, i.e., $s(\beta)=1$.
We assume the following input parameters :
%%%%%%%%%%%%%%%%%
\begin{eqnarray}
\beta = \sqrt{<\beta^{2}>} &=& 10^{-3}\;, \quad Z=1 \;, \quad  \adm = \alpha' = \alpha_{\mathrm{em}} \;,  \nonumber \\
 m_{X} &=& 1~\mathrm{GeV}\;, \quad
m_{\chi}=100~\mathrm{GeV} \;, \quad\kappa=1  \; . 
\label{4.52} 
\end{eqnarray}
%%%%%%%%%%%%%%%%%%%%%
%
Although this parameter space 
violates the bounds in  \eqs{enu}{bound-III} it serves as a benchmark
for this article in comparing results with those of section 3 if possible.
The value of $\kappa$ is chosen such that the resulting rate
presented in the figures assumes no mixing of the X-boson mediator
which is formally the case of model III.  

\begin{figure}[t]
   \centering
  \hbox{\includegraphics[height=2in]{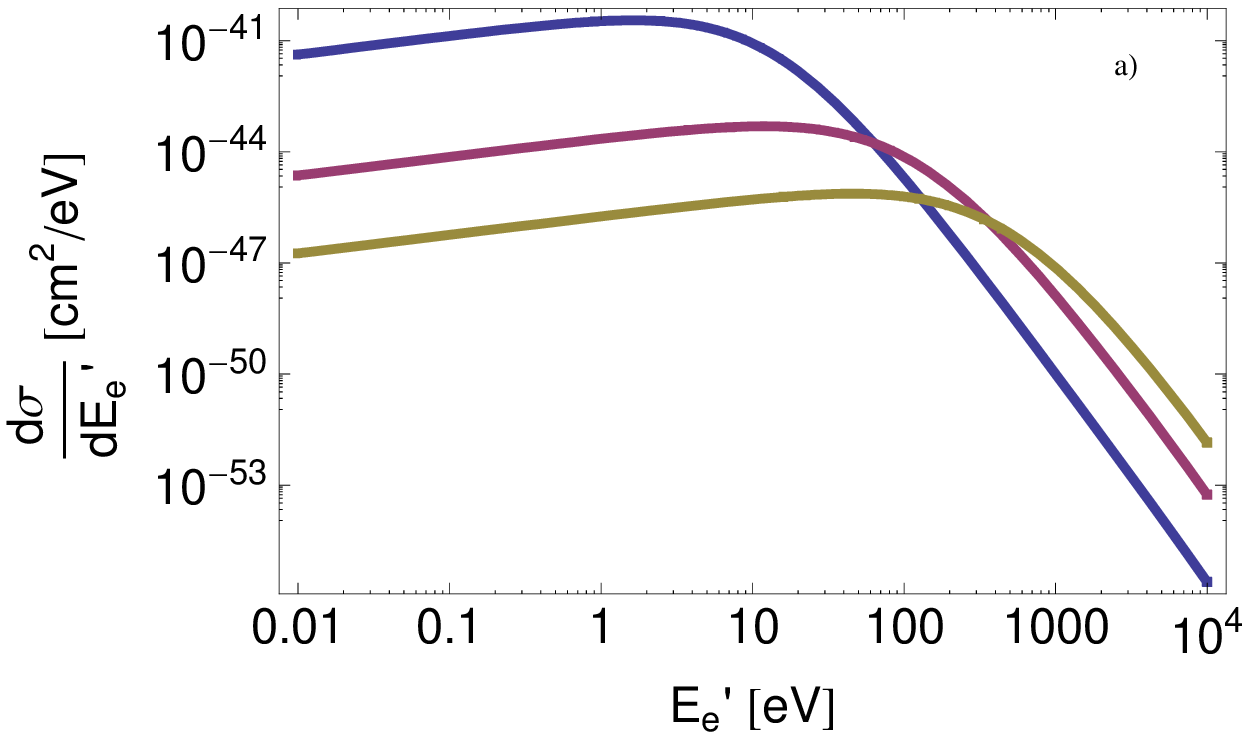}
   \hspace*{0.2cm}
   \includegraphics[height=1.9in]{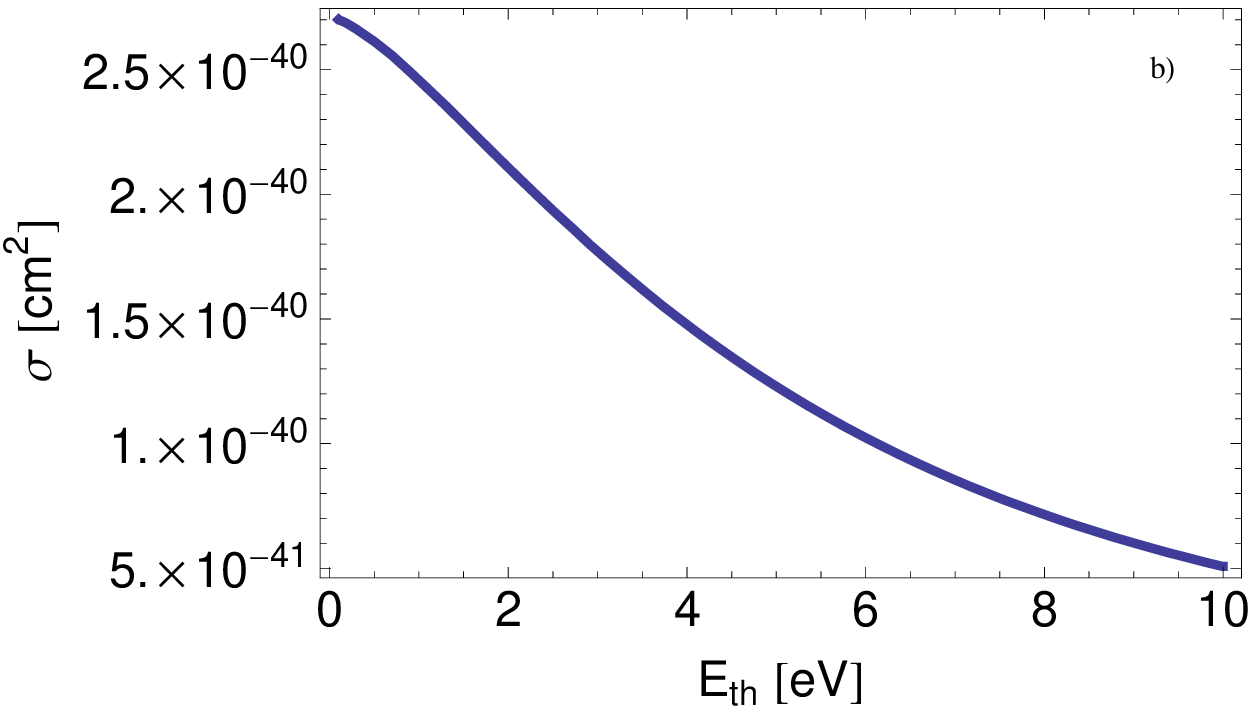}}
   \includegraphics[height=2in]{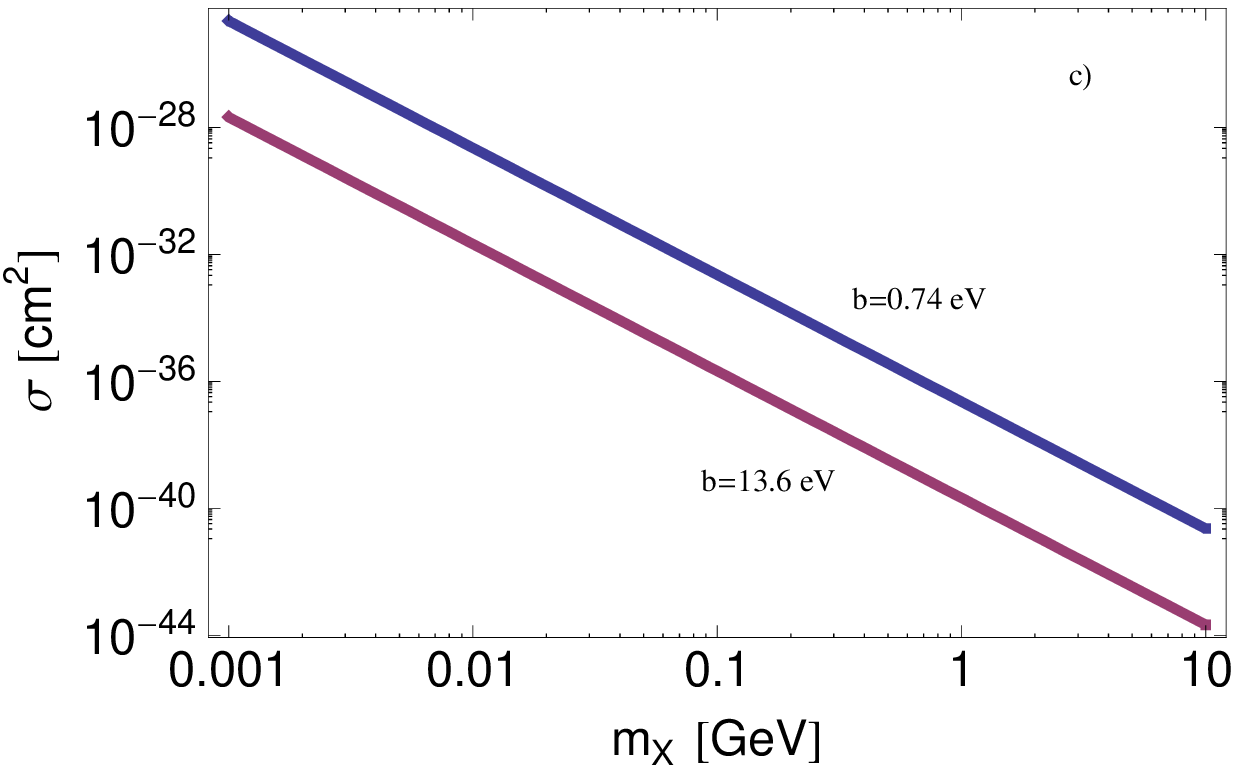}
   \caption{\sl a) Predictions for $d\sigma/dE_{e}'$ as a function of
   the ejected electron energy $E_{e}'$. The  target is assumed
   hydrogenic atom with  Z=1,3,6 (from top to bottom) in the ground state.
   b) The total cross section as a function of  threshold energy. c) The
   total cross section as a function of $m_{X}$ for two different 
   binding energies.
   We assume a Dirac WIMP, $E_{\rm th}=0$ eV and input parameters 
    from \eq{4.52} if not stated otherwise.}
       \label{fig:dsde}
\end{figure}

Results for the differential cross section $d\sigma / dE_{e}'$ for the
electron in the ground state of three hydrogenic atoms are
shown in Fig.~\ref{fig:dsde}a.
The differential cross section takes on its maximum values
 for final electron energy
 of around few eV for $Z=1$,  ten of eV for $Z=3$ and around
 hundred eV for $Z=6$. For the case $Z=1$, the extremum
  happens because of a fast increase of the
 term $\frac{|{\bf p_{\chi}'}|}{|{\bf p_{\chi}}|} \:
|{\bf p_{e}'}| \sim \sqrt{E_{e}'}$    and
the almost constant value of $|\phi_{100}|^{2}$
until 5 eV [see  \eq{eq1.20}]. For higher electron energies, e.g., $E_{e}' \gtrsim 10$ eV,  the
probability density factor $|\phi_{100}|^{2}$  drops
fast as $1/E^{'8}_{e}$ resulting in overall decreasing
of the cross section as $E_{e}^{'-15/2}$.
In physical terms, the outgoing electrons of high energy demand
high momenta   in the initial electron wavefunction, which leads
to suppression. The dependence on the $Z$ is easily explained 
if we recall that for hydrogenic atoms, 
$\langle p^{2} \rangle_{n=1} = Z^{2 }p_{0}^{2}$ where
$p_{0}$ is the Bohr momentum for Hydrogen.
Furthermore, 
despite appearances in \eq{eq1.20}, the differential cross section 
depends only very mildly on the  WIMP mass. One can show analytically
that the double integral over the wave function squared,  
is approximately proportional to
$1/m_{\chi}^{2}$ which cancels the $m_{\chi}^{2}$ in the numerator.

%%%%%%%%%%%%%%%%%%%%%%%%%%%%%%
%%%%%%%%%%%%%%%%%%%%%%%%%%%%%%%%%

%
Corresponding to the
input parameters noted in (\ref{4.52}) we calculate
the total cross section
from \eq{eq1.20} after numerical integration over $E_{e}'$
in the region $[E_{\rm th}, m_{\chi} \beta^{2}/2 - b(Z)]$.
For fixed velocity, $\beta = 0.001$, and $E_{\rm th}=0$ we find
the following representative values :
%%%%%%%%%%%%%%%%%%%%%
%%%%%%%%%%%%%%%%%%%%%%%
%\begin{table}[h]
\begin{center}
\begin{tabular}{c|c}$Z$ & $\sigma [\mathrm{cm}^{2}]$
 \\\hline
1 & $3 \times 10^{-40}$  \\\hline
10 & $2 \times 10^{-44}$  \\\hline
50 & $3 \times 10^{-48}$
\end{tabular}
%\caption{default}
\end{center}
\label{defaulttable}
%\end{table}
 %%%%%%%%%%%%%%%%%%%%%%%%%%%%
The total cross section
 increases by a factor of about $32$ when $\beta = \beta_{\rm esc}$
is taken. The cross section decreases with  
$Z$ [see also Fig.\ref{fig:dsde}a],   
the reason being the fact that the binding energy increases
with $Z^{2}$ [see  \eq{4.43}] and therefore we need to go
to larger - compared to ground state - momenta
where the wavefunction is small despite their maximum value
displacement towards larger momenta.

Assuming that the sensitivity of
detecting  low energy electrons will be analogous to the ongoing experiments ($\approx 10^{-43}~\mathrm{cm}^{2}$), we could even extract bounds on various parameters in models I, II or III. From
all running experiments, DAMA~\cite{DAMA,Bernabei}
 is the one that triggers on final
state electrons with energy around 5 KeV. From Fig.~\ref{fig:dsde}a
one obtains that, around that energy, the cross section is too small 
for $m_{X}=1$ GeV and all other inputs in \eq{4.52}.
However, $d\sigma/dE_{e}' \propto m_{X}^{-4}$ and therefore
for $m_{X} \approx 1$ MeV {\it i.e., } model types proposed 
in~\Ref{Boehm}, DAMA is a relevant experiment. 
Additionally, this
is  demonstrated  in  Fig.~\ref{fig:dsde}c where the total cross
section as a function of $m_{X}$ is plotted for two reference values
of binding energies. 

\begin{figure}[t]
   \centering
   \hbox{\includegraphics[height=2in]{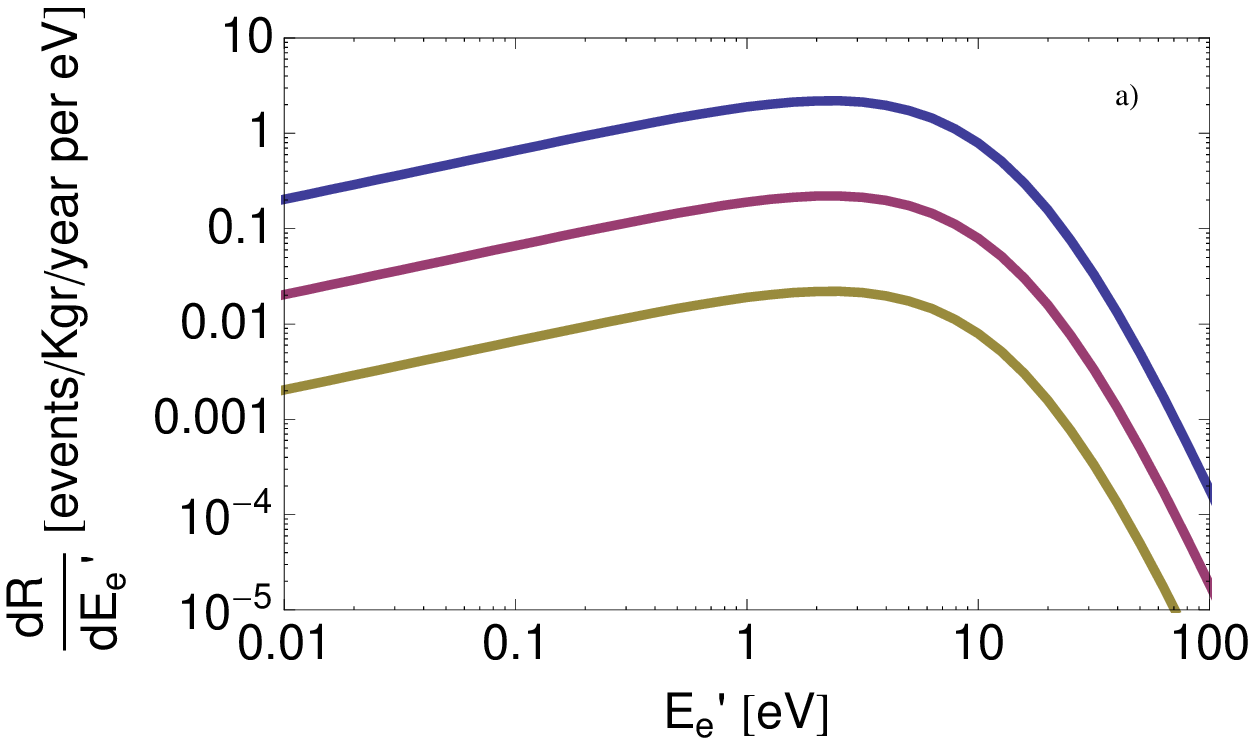}
   \hspace*{0.2cm}
   \includegraphics[height=2in]{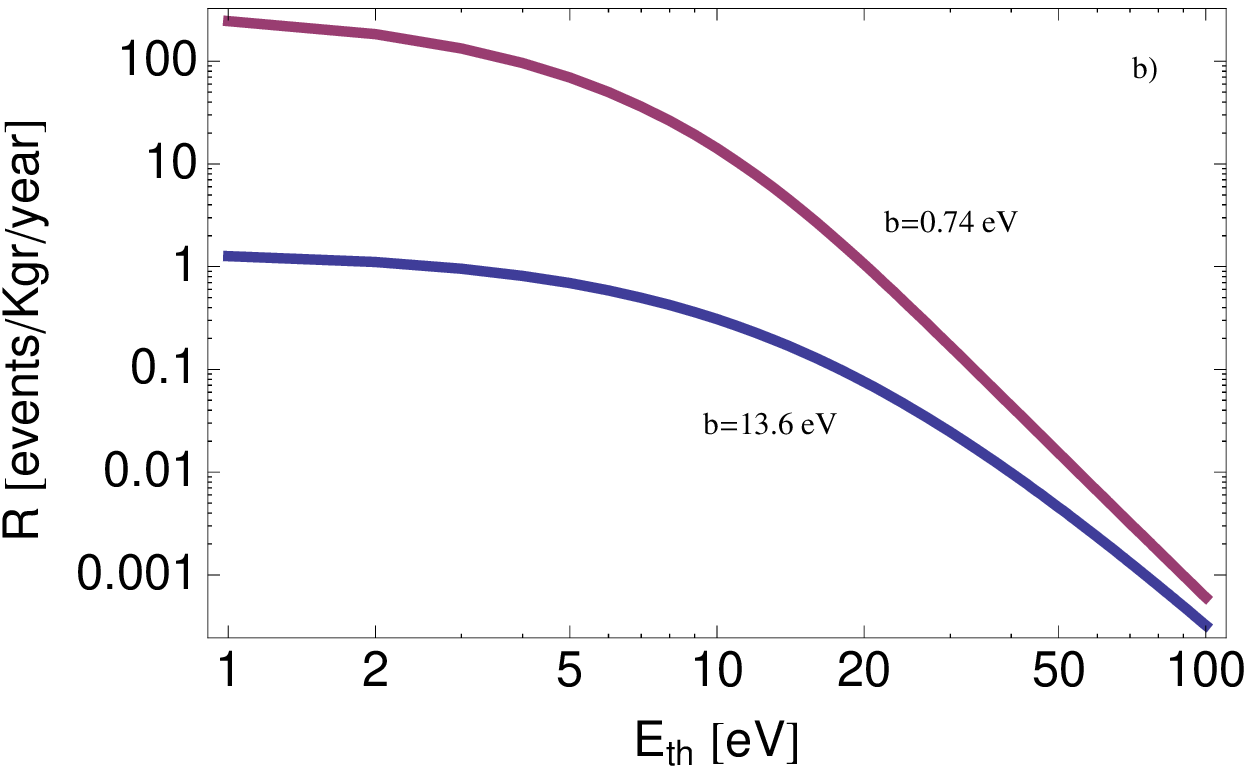}}
   \caption{\sl a) Differential event rate of Dirac WIMP
   scattered off hydrogen  (Z=1, A=1) target electrons
   per year per Kgr as a function of ejected electron energy $E_{e}'$ in eV.
   Three different WIMP masses have been assumed :
   $m_{\chi}   = 10, 100, 1000$ GeV, from top to bottom, respectively.
   b) The total event rate as a function of the experimental threshold
   energy for $m_{\chi}=100$ GeV for two different binding energies.
   Other input parameters are taken from \eq{4.52} for the
    massive mediator.
    }
   \label{fig:drde}
\end{figure}

In Fig.~\ref{fig:dsde}b we examine  the total
cross section as a function of the experimental energy threshold
for low energies, relevant to our proposal. As we can see,
the total
cross section reduces by a factor of six in the region
$0 \lesssim E_{\rm th} \lesssim 10$ eV. Above 10 eV the 
cross section drops drastically [see total rate in Fig.~\ref{fig:drde}b].

%
%
%%%%%%%%%%%%%%%%%%%%%%%%
%%%%%%%%%%%%%%%%%%%%%%%%%
%

%
Although not shown, 
we have also examined departures of the wavefunction
 from the ground state.
The maximum value $d\sigma/dE_{e}'|_{max}$ appears
at the same place in $E_{e}' \approx 1-10$ eV.
As an example, the difference in $d\sigma/dE_{e}'|_{max}$ is
an enhancement
by a factor 20 when going
from $1s \to 2s$.
Furthermore, the size of the  momentum transfer
in conjunction  with the non-zero binding energy are such that never let
the wavefunctions to reach their zero nodes.

Assuming one electron per target atom,
and the average cross section of Fig.~\ref{fig:dsde}a for $Z=1$,
the differential event rate per eV of electrons energy per year
per Kgr of hydrogen material as a functions of $E_{e}'$
for various WIMP masses
is depicted in Fig.~\ref{fig:drde}a.
The differential event rate again exhibits a maximum which
follows that of the differential cross section calculated in
Fig.~\ref{fig:dsde}a.
The event rate is of course higher for smaller WIMP mass [recall \eq{4.50}]
and for electron energy of few eV's it 
varies from 0.01 up to 2 events/yr/kgr/eV
for $m_{\chi}=1000,10$ GeV respectively.  For electron energy of around
100 eV the role of the wave function is to reduce the differential
rate by an order of magnitude 
i.e.,  from $10^{-4} \div 10^{-3}$ events/yr/kgr/eV. The total event
rate for  $m_{\chi}=100$ GeV and the other parameters in
\eq{4.52} is predicted to be:
\begin{equation}
R(Z=1,\kappa=1) \ \simeq \ 2~ \mathrm{[events/yr/target~kgr]} \;.
\label{eq:460}
\end{equation}
%%%%%%%%%%%%%%%%%%%%%%%%%%%
It is useful to know how the total rate (\ref{eq:460}) varies with
an experimental threshold energy. This information
can be extracted from Fig.~\ref{fig:drde}b for two different
but judiciously chosen,
values of binding energies. As in the case
of the total cross section in Fig.~\ref{fig:dsde}b, the total rate
drops by only a factor of five until $E_{\rm th} \approx 10$ eV 
while it drops very rapidly after about this scale. For example, it drops by a factor of $10^{4}$ for $E_{\rm th} = 100$ eV.  Smaller binding energies
[upper line in Fig.~\ref{fig:drde}b] result in up to two order of magnitude 
bigger rates but for threshold energies as low as
 $E_{\rm th} \lesssim 5$ eV.

Finally, in Table~2
we calculate the effects of time modulation
and present the
differential event rate for four different values of  $E_{e}'$ in 
 the case of massive mediator with  $m_{X}=1$ GeV. We assume
also a WIMP mass $m_{\chi}=100$ GeV and $Z=1$.
%%%%%%%%%%%%%%%%%%%%%
\begin{table}[t]
\begin{center}
\begin{tabular}{cccc}
\hline
\multicolumn{1}{c}{$E_{e}' $ [eV] }&
\multicolumn{2}{c}{$\left<\frac{dR}{dE_{e}'}\right>$ [events/kgr\,
target/year/eV]}\\
  & unmod. & mod.&H \\\hline
 %\hline
$0.1$ & $0.06$  & $0.01$& 0.17
\\ $1$ & $0.19$ & $0.02$&0.11
 \\ $10$ & $0.079$ & $0.008$&0.10
\\ $100$ & $1.84\times 10^{-5}$ & $1.78\times 10^{-6}$&0.097
\\\hline
\label{drdemodtable2}
\end{tabular}
\caption{\sl Time modulation effects  in case of a massive mediator
following \eq{tmmassless} and
various input parameters  in \eq{4.52}. $H$ 
 is the ratio of the modulated by the unmodulated 
 differential amplitude.}
\end{center}
\end{table}
 %%%%%%%%%%%%%%%%%%%%%%%%%%%%%%
The $H$ ratio  is constant around $10 \%$ independent of the energy and the WIMP mass. So the  modulation $h$ of the total rate is also going to be around $10 - 17\%$, which means that the difference between the maximum (here always in June 3nd) and the minimum (here  always in December) is  $20 - 34 \%$.

%%%%%%%%%%%%%%%%%%%%%%%%%%%%%%%%%
\subsection{Experiment :
The prospects of detecting single ultra low energy  electrons}

As discussed in a previous section observation of light X- boson would require detectors with sub-keV sensitivities. The development of such detectors, having a low energy threshold and low noise, remains generally a daunting challenge for present-day and future low- background experiments.

 As shown in Fig. 6 the signal of low energy electrons produced by elastic collision process exhibits a maximum at energies  around or even lower than 10 eV. At such energies a detector with single electron sensitivity will be required to reach a reasonable efficiency. 
The last ten years a particular effort is going on to develop ultra low threshold detectors  in order to address low energy neutrino 
physics~\cite{CG00,GV04,HB04,HTW08}. This has been  achieved for low mass detectors. We are, however, seeking an even lower energy threshold.

Usual solid state detectors employed for dark matter projects have typical thresholds of a few keV. It is very difficult to combine  sub-keV and big mass at the same time. For instance Ultra-Low-Energy Germanium detectors~\cite{CDMS} are able to reach a threshold of a few hundred eV's, but they are limited to a modular mass of a few grams. Anyway the achieved energy threshold is still below our requirements.

Single electron efficiency is achieved using detectors reaching very-high gains in order to cope with electronic noise. Gaseous detectors are good candidates. In such detectors high gains may easily be achieved. Having been conceived as a TPC Micromegas detector (µMS)~\cite{GIOM08}, it is compatible with large drift volumes and operation at high pressure, an example of which are the HELLAZ~\cite{Hellaz} prototypes. A great advantage of this detector is the versatility of target material: various gases from the lightest ($H_2$) to heaviest (Xe) could be used offering a large choice.

One idea to increase the mass of the target material is to use the recently developed Spherical Proportional Counter (SPC). This detector consists of large spherical gas volume with central electrode and radial electric field. Charges deposited in the drift volume are drifting to the central sensor where are amplified and collected. A novel concept of a proportional sensor, a metallic ball having a radius of about 15 mm, located at the center of curvature, acting as a proportional amplification structure is used. It allows to reach high gas gains ($\geq$ 104) and operates from low to high gas pressure. At such gains,  provided the low electronic noise of this detector,  single electron efficiency is easily achieved. 

The main advantages of the new structure relevant to our project are:
\begin{itemize}
\item	Simplicity of the design.
\item	A single channel is used to read-out a large volume.
\item	Robustness
\item	The depth of the interaction, related to the rise time of the signal, is measured. This is important to apply fiducial cuts for background rejection purpose.
\item	Low detector capacity $ \leq$ 0.1 pF, independent of the vessel size, allows very-low electronic noise, which is a key point toward  achieving low energy threshold.
\item	Versatility of the target material and density; the detector is compatible with a large variety of gases and could operate from low pressure to high pressure. This could be a precious tool to identify a possible signal out of backgrounds.
\end{itemize}

%%%%%%%%%%%%%%%%%%%%%%%%%%%%
A main concern of the proposed detection scheme is 
the minimal background level that will be reached by our system. 
By this one means that detector body and appropriate shield will 
be built with materials which are screened for low levels of 
natural and man-made radioactive impurities. Ordinary 
construction and shielding materials, however, do contain 
trace amounts of naturally occurring and man-made 
radionuclides which result in elevated background level; 
we need to design and fabricate the detector by careful 
material selection made out of low level activity.

Unfortunately, however, there
exists very little experience at
 the very low energy (sub keV) region where 
 our detector will be operating. An example 
 is a low background gaseous detector with sub KeV 
 energy threshold developed for solar axion search~\cite{Abbon};
  the reached background level is quite low and 
  is flat in the sub KeV energy range down to 250 eV. 
  Our purpose is to further decrease the energy threshold 
  down to about 10 eV. This region has never been 
  explored and therefore reaching the desired low level activity becomes a new experimental challenge. 
  Single electron backgrounds could be emitted
   by materials pulled by the electric field through 
   thermionic emission. The advantage of the spherical detector 
   is that at the external vessel the electric field is extremely low and therefore highly reduced thermionic emission is expected.

%%%%%%%%%%%%%%%%%%%%%%%%%%%%%

The present prototype having a volume of $1 m^{3}$, filled with a gas at high pressure with a target mass of the order of 10 kg could fulfills sensitivity requirements for our project.  We will search appropriate molecular gases having low binding energies and compatible with operation in the Spherical Proportional Counter detector~\cite{GIOMATARIS}.

At present it looks realistic to soon have a sphere of radius of 5 meters, which can be under a pressure of 5 bars. Thus, if one fills it with 80\% Ar and 20\% Isobutane ($C_4H_{10}$), one can have 212 Kg of Hydrogen.  With this much Hydrogen using \eq{eq:460} and a threshold of $\approx$10 eV,  we expect around 
200 events per year for the parameters in (\ref{4.52}). 
In models~\cite{Boehm}
where the mediator mass is very low, 
{\it e.g.} $m_{X }\approx 1$ MeV, we expect
an increase of the event rate by almost six orders of magnitude. 
Therefore, if a low energy experiment will be  
built it would possibly set the best limits 
on these kind of models.

\section{Conclusions}

Recent cosmic ray results from PAMELA, HESS and FERMI
collaborations  show an unexpected rising of positron events
with energy
that may be due to Dark Matter particle annihilations in the halo of our
Galaxy.   This Dark Matter particle ``sees'' the SM ones only
through its interactions with an X-boson that couples to 
the SM gauge sector. Depending on the model, the mediator 
can be massless  or massive with different couplings. In this article
 we study direct detection of this secluded type of dark matter employing
 nucleons or electrons with main emphasis in the latter case.
 
 Due to the small momentum transfer\footnote{For nucleons, the momentum transfer is $\approx 2$ MeV and energy transfer 
 is $\approx 2$ KeV, while for low energy  electron recoils 
 they are  $\approx 50$ KeV  and  
 $\approx 10$ eV, respectively.} the massless case
 results in a large number of events that should have been seen by
 current nucleon recoiling direct detection experiments and therefore
 strong bounds on mixing parameters and couplings exist.
 Our work emphasizes the role of the low energy electron recoil 
 in direct detection experiments and proposes a novel 
 experimental avenue on how to proceed in searching for
 such low energy electrons. For  simple
 hydrogenic atoms, and  at low energy, $E_{e}' \approx 10$ eV,
 the cross section is enhanced by order of magnitudes compared
 to KeV recoil energies. In the neighborhood of low energies, the 
 results depend highly on the binding energy of the ejected electron:
 the more loose the electron is the bigger the event rate becomes as
 expected. 
 In this regard we considered two possibilities:
\begin{itemize}
\item The process is mediated by the massive mediator X (our model III).\\
In this case we do not have scattering off hadrons at tree level. So we do not have dominant constraints on the parameters of the model coming from the ongoing WIMP searches. Using the parameters of \eq{4.52} we have obtained fairly large cross sections for a Dirac WIMP. Employing the spherical TPC detector described above with a radius of 5 $m$ under pressure of 5 Atm we have found that we could have about 200 counts in a year, assuming a threshold of 10 eV. It is possible, however, that our choice of parameters is a bit optimistic and we may have not considered all available constraints.
Our results are also applicable to model-I. In this case however,
due to the fact that couplings of the X-boson to hadrons appear
at tree level, there exist strong
constraints on the mixing parameter already from the nucleon direct searches [see \eq{nc2}] .
\item The process is accommodated by the massless mediator 
(leptophylic version of model II)\\
This mechanism is similar to that involving hadrons in section~3, 
one simply replaces the quarks by leptons. In this case we have 
found that the most stringent constraints on 
the parameters come from the standard WIMP searches. 
Thus using the parameters of Eq. (4.65) we 
have obtained with the above detector hundreds of events per year
even with a (reduction) mixing coupling 
constant as low as $\kappa = 10^{-10}$ for a Dirac WIMP. 
{\it Such a huge signal cannot be seen by current experiments
either due to lack of low energy threshold or because,  experiments,
like CDMS and XENON, are keeping only nuclear recoil events.}
We were surprised to find so large cross section.
 We now understand it, however, to be due 
 to the photon propagator $(1/q^2)^2$, 
 which is favored by the fact that the momentum transfer 
 is very low in the case of electrons. 
 We should mention that, since the initial electron is bound, 
 there is no infrared divergence and no 
 need for a low energy cut off. It should be also
noted that quark couplings to X-boson 
will come back through loop corrections
even if they are  forbidden
at tree level by a symmetry which is eventually broken [see footnote 2].
Then current nucleon recoil experiments will be as important
[see \eq{nc1}] and complementary to the electron ones.  
\end{itemize}
The above conclusions  assume that the WIMP is a Dirac particle.
If the WIMP is Majorana particle the rates are suppressed by
approximately a factor  $\beta^{2} \approx 10^{-6}$. For both
the above cases, annual time modulation effects are of the 
order of 20-30\%, important enough to be noticed.

We have limited the discussion of the rates in the case
 of hydrogen, since our cross section was evaluated using
   hydrogenic wave functions. Certainly the 
obtained rates will increase, if one can exploit the  
other atomic electrons with smaller binding energy.
This situation was made manifest in our work with 
a judicious change of the binding energy [see Figs.~6b,7c,8b].
But then one should  employ realistic wave functions.

In a similar fashion one can treat other dark matter candidates like right handed neutrinos, which arise in models in which the ordinary Dirac type mass is forbidden due to a discreet symmetry, but communication with the leptons is allowed via exotic scalars~\cite{AEM08,Ma2,Yoshida} with masses in the 50 GeV region. It may also apply to other models involving exotic fermions and scalars recently proposed and reviewed in
 \Ref{MA08}.

\medskip

\subsection*{Acknowledgements}

A.D., and J.D.V.,  acknowledge
partial support  by the EU FP6 Marie
Curie Research and Training Network ``UniverseNet" (MRTN-CT-2006-035863).
A.D. is also partially supported by the RTN European Programme,
MRTN-CT-2006-035505 (HEPTOOLS, Tools and Precision Calculations
for Physics Discoveries at Colliders).
K.S. acknowledges full financial support from Greek State
Scholarships Foundation (I.K.Y).
%%%%%%%%%%%%%%%%%%%%%%%%%%%%%%%%


\bigskip

\begin{thebibliography}{99}


\bibitem{Pamela1}
  O.~Adriani {\it et al.},
  %``Observation of an anomalous positron abundance in the cosmic radiation,''
  arXiv:0810.4995 [astro-ph].
  %%CITATION = ARXIV:0810.4995;%%

\bibitem{Pamela2}
  O.~Adriani {\it et al.},
  %``A new measurement of the antiproton-to-proton flux ratio up to 100 GeV in
  %the cosmic radiation,''
  arXiv:0810.4994 [astro-ph].
  %%CITATION = ARXIV:0810.4994;%%


\bibitem{HEAT1}
S.W. Barwick {\it et al}, Astrophys. J. {\bf 482} (1997) l191; astro-ph/9703192.

\bibitem{HEAT2}
J.J. Beatty {\it et al}, Phys. Rev. Let. {\bf 93} (2004) 241102; astro-ph/0412230.

\bibitem{AMS01}
  M.~Aguilar {\it et al.}  [AMS-01 Collaboration],
  %``Cosmic-ray positron fraction measurement from 1-GeV to 30-GeV with
  %AMS-01,''
  Phys.\ Lett.\  B {\bf 646} (2007) 145
  [arXiv:astro-ph/0703154].
  %%CITATION = PHLTA,B646,145;%%

\bibitem{FERMI}
  A.~A.~Abdo {\it et al.}  [The Fermi LAT Collaboration],
  %``Measurement of the Cosmic Ray e+ plus e- spectrum from 20 GeV to 1 TeV with
  %the Fermi Large Area Telescope,''
  arXiv:0905.0025 [astro-ph.HE].
  %%CITATION = ARXIV:0905.0025;%%

\bibitem{HESS}
 F.~Aharonian {\it et al.}  [H.E.S.S. Collaboration],
  %``The energy spectrum of cosmic-ray electrons at TeV energies,''
  Phys.\ Rev.\ Lett.\  {\bf 101} (2008) 261104
  [arXiv:0811.3894 [astro-ph]];
  %%CITATION = PRLTA,101,261104;%%
  H.~E.~S.~Aharonian,
  %``Probing the ATIC peak in the cosmic-ray electron spectrum with H.E.S.S,''
  arXiv:0905.0105 [astro-ph.HE].
  %%CITATION = ARXIV:0905.0105;%%

\bibitem{Strumia}
 M.~Cirelli, M.~Kadastik, M.~Raidal and A.~Strumia,
  %``Model-independent implications of the e+, e-, anti-proton cosmic ray
  %spectra on properties of Dark Matter,''
  Nucl.\ Phys.\  B {\bf 813}, 1 (2009)
  [arXiv:0809.2409 [hep-ph]].
  %%CITATION = NUPHA,B813,1;%%
  
\bibitem{Meade}
  P.~Meade, M.~Papucci, A.~Strumia and T.~Volansky,
  %``Dark Matter Interpretations of the Electron/Positron Excesses after
  %FERMI,''
  arXiv:0905.0480 [hep-ph].
  %%CITATION = ARXIV:0905.0480;%%



\bibitem{ATIC}
J.~Chang, {\it et al}, [ATIC Collaboration], Nature, 456, 362 (2008).


\bibitem{WMAP1}
D. P. Finkbeiner {\it et al}, Astrophys. J. {\bf 684} (2004) 186; astro-ph/0312547.

\bibitem{WMAP2}
D. Hooper, D. P. Finkbeiner and G. Dobler , Phys. Rev. D {\bf 76} (2007) arXiv 0705.3655.

\bibitem{EGRET}
A.W. Strong {\it et al}, Astron. Astrophys. {\bf 444} (2005) 405; astro-ph/0509092.

\bibitem{Holdom}
B.~Holdom, Phys. \ Lett. \ B {\bf 166}, 196 (1986);
{\bf \it ibid.} B {\bf 259} 329 (1991).


\bibitem{Maxim}
  M.~Pospelov, A.~Ritz and M.~B.~Voloshin,
  %``Secluded WIMP Dark Matter,''
  Phys.\ Lett.\  B {\bf 662}, 53 (2008)
  [arXiv:0711.4866 [hep-ph]].
  %%CITATION = PHLTA,B662,53;%%
Kinetic mixing of the photon with hidden $U(1)$s 
in string theory has been studied in 
S.~A.~Abel, M.~D.~Goodsell, J.~Jaeckel, V.~V.~Khoze and A.~Ringwald,
  %``Kinetic Mixing of the Photon with Hidden U(1)s in String Phenomenology,''
  JHEP {\bf 0807}, 124 (2008)
  [arXiv:0803.1449 [hep-ph]].
  %%CITATION = JHEPA,0807,124;%%

\bibitem{Finkbeiner}
D.~P.~Finkbeiner and N.~Weiner,
  %``Exciting Dark Matter and the INTEGRAL/SPI 511 keV signal,''
  Phys.\ Rev.\  D {\bf 76}, 083519 (2007)
  [arXiv:astro-ph/0702587].
  %%CITATION = PHRVA,D76,083519;%%


\bibitem{models1}
N.~Arkani-Hamed, D.~P.~Finkbeiner, T.~R.~Slatyer and N.~Weiner,
  %``A Theory of Dark Matter,''
  Phys.\ Rev.\  D {\bf 79} (2009) 015014
  [arXiv:0810.0713 [hep-ph]];
  %%CITATION = PHRVA,D79,015014;%%
N.~Arkani-Hamed and N.~Weiner,
  %``LHC Signals for a SuperUnified Theory of Dark Matter,''
  JHEP {\bf 0812} (2008) 104
  [arXiv:0810.0714 [hep-ph]];
  %%CITATION = JHEPA,0812,104;%%
M.~Pospelov and A.~Ritz,
  %``Astrophysical Signatures of Secluded Dark Matter,''
  arXiv:0810.1502 [hep-ph].
  %%CITATION = ARXIV:0810.1502;%%




\bibitem{Sommerfeld}
 A.~Sommerfeld, Ann. \ Phys. \ {\bf 11} 257 (1931);
 
 Relevant to the abelian models considered here is the
 article, M.~Cirelli, A.~Strumia, M.~Tamburini, Nucl. \ Phys. \
 B {\bf 787} (2007) [arXiv:0706.4071 [hep-ph]].


\bibitem{Boehm}
C.~Boehm and P.~Fayet,
  %``Scalar dark matter candidates,''
  Nucl.\ Phys.\  B {\bf 683} (2004) 219
  [arXiv:hep-ph/0305261];
  %%CITATION = NUPHA,B683,219;%%
  C.~Boehm, P.~Fayet and J.~Silk, Phys. \ Rev. \ D {\bf 69},
  101302 (2004) [arXiv:hep-ph/0311143].


\bibitem{Stueckelberg}
  E.~C.~G.~Stueckelberg,
  %``Interaction energy in electrodynamics and in the field theory of nuclear
  %forces,''
  Helv.\ Phys.\ Acta {\bf 11}, 225 (1938).
  %%CITATION = HPACA,11,225;%%


\bibitem{Nath2}
  D.~Feldman, Z.~Liu and P.~Nath,
  %``PAMELA Positron Excess as a Signal from the Hidden Sector,''
  arXiv:0810.5762 [hep-ph].
  %%CITATION = ARXIV:0810.5762;%%




\bibitem{Fox}
  P.~J.~Fox and E.~Poppitz,
  %``Leptophilic Dark Matter,''
  arXiv:0811.0399 [hep-ph].
  %%CITATION = ARXIV:0811.0399;%%

\bibitem{Baek}
  S.~Baek and P.~Ko,
  %``Phenomenology of $U(1)_{L_\mu - L_\tau}$ charged dark matter at PAMELA and
  %colliders,''
  arXiv:0811.1646 [hep-ph].
  %%CITATION = ARXIV:0811.1646;%%

\bibitem{Kribs}
 R.~Harnik and G.~D.~Kribs,
  %``An Effective Theory of Dirac Dark Matter,''
  arXiv:0810.5557 [hep-ph].
  %%CITATION = ARXIV:0810.5557;%%

\bibitem{Yanagida}
C.~R.~Chen, F.~Takahashi and T.~T.~Yanagida,
  %``High-energy Cosmic-Ray Positrons from Hidden-Gauge-Boson Dark Matter,''
  arXiv:0811.0477 [hep-ph].
  %%CITATION = ARXIV:0811.0477;%%

\bibitem{Ringwald}
A.~Ibarra, A.~Ringwald, D.~Tran and C.~Weniger,
  %``Cosmic Rays from Leptophilic Dark Matter Decay via Kinetic Mixing,''
  JCAP {\bf 0908} (2009) 017
  [arXiv:0903.3625 [hep-ph]].
  %%CITATION = JCAPA,0908,017;%%

\bibitem{Bernabei}
R.~Bernabei {\it et al.},
  %``Investigating electron interacting dark matter,''
  Phys.\ Rev.\  D {\bf 77}, 023506 (2008)
  [arXiv:0712.0562 [astro-ph]].
  %%CITATION = PHRVA,D77,023506;%%
  
In the case of inelastic dark matter and DAMA prospects
see, Y.~Cui, D.~E.~Morrissey, D.~Poland and L.~Randall,
  %``Candidates for Inelastic Dark Matter,''
  JHEP {\bf 0905}, 076 (2009)
  [arXiv:0901.0557 [hep-ph]].
  %%CITATION = JHEPA,0905,076;%%



\bibitem{Sakurai}
P.~Q.~Hung and J.~J.~Sakurai,
  %``Gamma W0 Mixing As An Alternative To Unified Weak Electromagnetic Gauge
  %Theories,''
  Nucl.\ Phys.\  B {\bf 143} (1978) 81
  [Erratum-ibid.\  B {\bf 148} (1979) 538].
  %%CITATION = NUPHA,B143,81;%%


\bibitem{Cheung}
  M.~Baumgart, C.~Cheung, J.~T.~Ruderman, L.~T.~Wang and I.~Yavin,
  %``Non-Abelian Dark Sectors and Their Collider Signatures,''
  arXiv:0901.0283 [hep-ph].
  %%CITATION = ARXIV:0901.0283;%%



\bibitem{HAGI}
For a recent review see, F.~Jegerlehner and A.~Nyffeler,
  %``The Muon g-2,''
  arXiv:0902.3360 [hep-ph].
  %%CITATION = ARXIV:0902.3360;%%

\bibitem{Maxim2}
  M.~Pospelov,
  %``Secluded U(1) below the weak scale,''
  arXiv:0811.1030 [hep-ph].
  %%CITATION = ARXIV:0811.1030;%%


\bibitem{Zurek}
  D.~E.~Morrissey, D.~Poland and K.~M.~Zurek,
  %``Abelian Hidden Sectors at a GeV,''
  arXiv:0904.2567 [hep-ph].
  %%CITATION = ARXIV:0904.2567;%%

\bibitem{Bjorken}
J.~D.~Bjorken, R.~Essig, P.~Schuster and N.~Toro,
  %``New Fixed-Target Experiments to Search for Dark Gauge Forces,''
  arXiv:0906.0580 [hep-ph].
  %%CITATION = ARXIV:0906.0580;%%




\bibitem{Maxim3}
B.~Batell, M.~Pospelov and A.~Ritz,
  %``Exploring Portals to a Hidden Sector Through Fixed Targets,''
  arXiv:0906.5614 [hep-ph].
  %%CITATION = ARXIV:0906.5614;%%


\bibitem{Nath}
 B.~Kors and P.~Nath,
  %``A Stueckelberg extension of the standard model,''
  Phys.\ Lett.\  B {\bf 586} (2004) 366
  [arXiv:hep-ph/0402047];
  %%CITATION = PHLTA,B586,366;%%
  D.~Feldman, Z.~Liu and P.~Nath,
  %``The Stueckelberg Z' extension with kinetic mixing and milli-charged dark
  %matter from the hidden sector,''
  Phys.\ Rev.\  D {\bf 75} (2007) 115001
  [arXiv:hep-ph/0702123].
  %%CITATION = PHRVA,D75,115001;%%
 
\bibitem{CDMS}
  J.~Yoo  [CDMS Collaboration],
  %``Results from the CDMS 5-Tower Operation,''
  arXiv:0810.3527 [hep-ex].
  %%CITATION = ARXIV:0810.3527;%%

\bibitem{XENON}
J.~Angle {\it et al.}  [XENON Collaboration],
  %``First Results from the XENON10 Dark Matter Experiment at the Gran Sasso
  %National Laboratory,''
  Phys.\ Rev.\ Lett.\  {\bf 100} (2008) 021303
  [arXiv:0706.0039 [astro-ph]].
  %%CITATION = PRLTA,100,021303;%%



 \bibitem{DAMA}
R.~Bernabei {\it et al.}  [DAMA Collaboration],
  %``First results from DAMA/LIBRA and the combined results with DAMA/NaI,''
  Eur.\ Phys.\ J.\  C {\bf 56}, 333 (2008)
  [arXiv:0804.2741 [astro-ph]].
  %%CITATION = EPHJA,C56,333;%%
 
\bibitem{Vergados}
J.~D.~Vergados,
  %``Direct SUSY dark matter detection: Theoretical rates due to the spin,''
  J.\ Phys.\ G {\bf 30}, 1127 (2004)
  [arXiv:hep-ph/0406134].
  %%CITATION = JPHGB,G30,1127;%%
  
\bibitem{TETRVER06}
N.Tetradis, J.D. Vergados and A. Faessler, Phys. Rev. D {\bf 75}, 023504 (2007)


\bibitem{Peskin}
See for instance,
  M.~E.~Peskin and D.~V.~Schroeder,
  ``An Introduction To Quantum Field Theory,''
{\it  Reading, USA: Addison-Wesley (1995) 842 p}.

\bibitem{Schiff}
See for example, 
%

Hans~A.~Bethe, ``Intermediate Quantum Mechanics,''
{\it Lecture notes and supplements in physics}. Notes by
R.~W.~Jackiw. 
%

L. Schiff,
``Quantum Mechanics,''
{\it McGraw-Hill Education (1968) 584p}.


% %%%%%%   EXP SECTION  %%%%%%%%
\bibitem{CG00}
	J. I Collar, I. Giomataris, 
	Nucl. \ Instrum. \ Meth. \ {\bf A471}:254-259,2000.

\bibitem{GV04}
	I. Giomataris, J, D. Vergados, 
	Nucl. \ Instrum. \ Meth. \ {\bf A530}:330-358,2004.

\bibitem{HB04}
	Chris Hagmann, Adam Bernstein, 
	IEEE Trans. \ Nucl. \ Sci.\ {\bf 51}:2151-2155,2004.

\bibitem{HTW08}
	Henry T. Wong, Mod. \ Phys. \ Lett.\ {\bf A23}:1431-1442,2008.




\bibitem{GIOM08}
Y.~Giomataris, P.~Rebourgeard, J.~P.~Robert and G.~Charpak,
  %``MICROMEGAS: A high-granularity position-sensitive gaseous detector for
  %high particle-flux environments,''
  Nucl.\ Instrum.\ Meth.\  A {\bf 376} (1996) 29.
 %%CITATION = NUIMA,A376,29;%%

\bibitem{Hellaz}
P.~Gorodetzky {\it et al.},
  %``Identification of solar neutrinos by individual electron counting in
  %HELLAZ,''
  Nucl.\ Instrum.\ Meth.\  A {\bf 433}, 554 (1999).
  %%CITATION = NUIMA,A433,554;%%


\bibitem{SiPM}
P.~Buzhan {\it et al.},
  %``Silicon photomultiplier and its possible applications,''
  Nucl.\ Instrum.\ Meth.\  A {\bf 504}, 48 (2003).
  %%CITATION = NUIMA,A504,48;%%

\bibitem{Abbon}
P.~Abbon {\it et al.},
  %``The Micromegas detector of the CAST experiment,''
  New J.\ Phys.\  {\bf 9} (2007) 170
  [arXiv:physics/0702190].
  %%CITATION = NJOPF,9,170;%%


\bibitem{GIOMATARIS}
I.~Giomataris {\it et al.},
  %``A novel large-volume Spherical Detector with Proportional Amplification
  %read-out,''
  JINST {\bf 3}, P09007 (2008)
  [arXiv:0807.2802 [physics.ins-det]].
  %%CITATION = JINST,3,P09007;%%




\bibitem{AEM08}
R.~Adhikari, J.~Erler and E.~Ma,
  %``Seesaw Neutrino Mass and New U(1) Gauge Symmetry,''
  Phys.\ Lett.\  B {\bf 672} (2009) 136
  [arXiv:0810.5547 [hep-ph]].
  %%CITATION = PHLTA,B672,136;%%

\bibitem{Ma2}
Q.~H.~Cao, E.~Ma and G.~Shaughnessy,
  %``Dark Matter: The Leptonic Connection,''
  arXiv:0901.1334 [hep-ph].
  %%CITATION = ARXIV:0901.1334;%%

\bibitem{Yoshida}
D.~Suematsu, T.~Toma and T.~Yoshida,
  %``Reconciliation of CDM abundance and $\mu\to e\gamma$ in a radiative seesaw
  %model,''
  arXiv:0903.0287 [hep-ph].
  %%CITATION = ARXIV:0903.0287;%%

\bibitem{MA08}
E.~Ma,
  %``Neutrino Mass Seesaw Version 3: Recent Developments,''
  arXiv:0810.5574 [hep-ph].
  %%CITATION = ARXIV:0810.5574;%%




\end{thebibliography}
\end{document}